\documentclass[aps,pra,reprint,superscriptaddress]{revtex4-1}
\usepackage{amssymb}
\usepackage{amsmath}
\usepackage{graphicx}
\usepackage{epstopdf}
\usepackage{subfigure}
\newcommand{\tabincell}[2]{
\begin{tabular}{@{}#1@{}}#2\end{tabular}
}
\begin{document}
\title{Noise Analysis for High-Fidelity Quantum Entangling Gates in an Anharmonic Linear Paul Trap}
\author{Yukai Wu}
\affiliation{Department of Physics, University of Michigan, Ann Arbor, Michigan 48109, USA}
\author{Sheng-Tao Wang}
\affiliation{Department of Physics, University of Michigan, Ann Arbor, Michigan 48109, USA}
\affiliation{Department of Physics, Harvard University, Cambridge, Massachusetts 02138, USA}
\author{L.-M. Duan}
\affiliation{Department of Physics, University of Michigan, Ann Arbor, Michigan 48109, USA}
\affiliation{Center for Quantum Information, IIIS, Tsinghua University, Beijing 100084, P. R. China}
\date{\today}

\begin{abstract}
The realization of high fidelity quantum gates in a multi-qubit system, with a typical target set at $99.9\%$, is a critical requirement for the implementation of fault-tolerant quantum computation. To reach this level of fidelity, one needs to carefully analyze the noises and imperfections in the experimental system and optimize the gate operations to mitigate their effects. Here, we consider one of the leading experimental systems for the fault-tolerant quantum computation, ions in an anharmonic linear Paul trap, and optimize entangling quantum gates using segmented laser pulses with the assistance of all the collective transverse phonon modes of the ion crystal. We present detailed analyses of the effects of various kinds of intrinsic experimental noises as well as errors from imperfect experimental controls. Through explicit calculations, we find the requirements on these relevant noise levels and control precisions to achieve the targeted high fidelity of $99.9\%$ for the entangling quantum gates in a multi-ion crystal.
\end{abstract}

\maketitle

\section{Introduction}
The trapped ion system is one of the most promising candidates for large-scale quantum computing because of its long coherence time, nearly perfect initialization and detection methods, and also the strong laser-mediated Coulomb interaction among ions which facilitates long-range entangling gates \cite{Cirac1995,Blatt2008,Monroe2013}. Many pioneering works have laid the building block of a scalable ion-trap quantum computer \cite{Blatt2008,Monroe2013,ladd2010quantum}. For example, recently, fidelity higher than 99.99\% for a single-qubit gate and 99.9\% for an entangling gate in a two-ion crystal have been reported \cite{Ballance2016,Gaebler2016}. Meanwhile, some important quantum algorithms, such as Shor's algorithm and quantum error correction, have been demonstrated in small scale \cite{Monz2016,Chiaverini2004,Schindler2011}.

One remaining problem is how to scale up the system. For a small number of ions, one scheme to realize the entangling gate, known as the Molmer-Sorensen (MS) gate, has been proposed for almost two decades~\cite{Sorensen1999}. It utilizes a single phonon mode of the ion crystal, typically the center-of-mass mode, to mediate a coupling between two ions' internal states, which is insensitive to the phonon number. However, as the number of ions increases, the motion of the ion crystal becomes progressively more complex and the crosstalk among different collective modes can lead to errors in the quantum gate \cite{Blatt2008}. A straightforward approach to suppress this crosstalk is to weaken the laser driving, but at the cost of increasing the gate time with the number of ions. One possible solution is to use an architecture called the quantum charge-coupled device \cite{Kielpinski2002,Hensinger2006,Monroe2013}, where entanglements are first generated in individual zones and are then distributed to other regions by a classical ion shuttling technique. Such a shuttling, however, demands exquisite control of ion positions. In this paper we will focus on a different approach, where all the collective modes are utilized to perform optimized entangling gates \cite{Zhu2006}. In this way, the existence of multiple phonon modes is no longer a source of error. One can then use segmented laser pulses to optimize the gate performance.

Many theoretical and experimental works have been done along this path \cite{Zhu2006PRL-transverse-mode,Lin2009,Choi2014,Wang2015}. For example, it was proposed that a suitable quartic potential can be applied to make the ion spacing of a linear chain more uniform~\cite{Lin2009}. In this setup, the ion structure will be more stable against the zigzag shape and it has a narrower transverse phonon spectrum, allowing more efficient cooling and control.

The original scheme of Ref.~\cite{Zhu2006} uses many approximations. However, a detailed and systematic error analysis, which is essential for high gate fidelity above the fault-tolerant error threshold, is still lacking. On the other hand, such an error analysis has been made in Ref.~\cite{Ballance2017} for a two-ion crystal, but its scheme and the numerical methods cannot be directly applied to a multi-ion crystal. In this paper, we thoroughly examine the approximations made in the scalable scheme and the influence of fluctuation in gate parameters. Many of our analyses can also be applied to other protocols based on extensions of the MS gate. Here we focus on a one-dimensional (1D) ion crystal, while a generalization to other structures is straightforward.

The paper is organized in the following way. In Sec.~\ref{sec:method} we review the scheme to realize the XX entangling gate between any pair of ions in a large ion crystal. Then in Sec.~\ref{sec:result} an example is presented for a chain of 19 ${}^{171}\mathrm{Yb}^+$ ions. We choose this size of the ion crystal since it is the size of the current experimental platform for the demonstration of a logic qubit. The formalism and many of the analyses in this paper are directly applicable to ion crystals of any size. We numerically optimize the gate parameters to realize high-fidelity entangling gates between ions with different separations. Stability of the gate under fluctuation in these parameters is then discussed, together with a comprehensive list of the source of errors from the approximations and the neglected physical effects. We then conclude in Sec.~\ref{sec:conclusion}. Appendix~\ref{app:anharmonic} describes the technique to find the ions' equilibrium configuration and the collective phonon modes. Appendix~\ref{app:higher_order} discusses in depth the effects of higher order terms neglected in our formulation. Then we examine a key consideration in our scheme, the asymmetry in laser beams, in Appendix~\ref{app:asymmetry}. Finally in Appendix~\ref{app:accumulation} we discuss how the errors from imperfect gate design accumulate when the gates are applied repeatedly.
\section{Entangling Gates in a Linear Paul Trap}
\label{sec:method}
\subsection{Hamiltonian and Time Evolution Operator}
Consider $N$ ions in a linear Paul trap along the $z$ axis. A suitable quartic potential can be applied in the $z$ direction through external electrodes, making the spacings of ions nearly uniform. For typical experimental parameters, the micro-motion is small and can be neglected. Then we can calculate the equilibrium configuration as well as the collective oscillation modes. See Appendix~\ref{app:anharmonic} for more details about the derivation.

\begin{figure}[!tbp]
  \centering
  \includegraphics[width=0.8\linewidth]{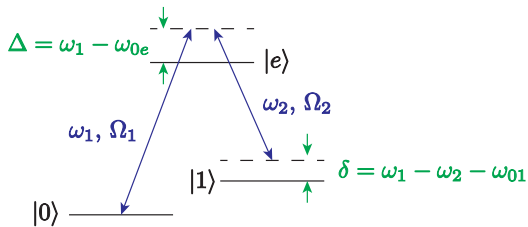}\\
  \caption{Schematic diagram of the three levels and the two Raman beams.}\label{fig:Raman}
\end{figure}

We start from a three-level approximation of the ion's level structure (Fig.~\ref{fig:Raman}). Later we will adiabatically eliminate the excited state to attain the two-level approximation. The free Hamiltonian of this system is
\begin{equation}
H = \hbar\sum_{i=1}^N \left( \omega_{01} |1\rangle_i\langle 1| + \omega_{0e} |e\rangle_i\langle e| \right) + \hbar\sum_k \omega_k a_k^\dag a_k,
\end{equation}
where $\hbar\omega_{01}\equiv E_1 - E_0$ is the energy difference between $|0\rangle$ and $|1\rangle$ (typically two hyperfine ``clock'' states of the ion \cite{Choi2014}) and $\hbar\omega_{0e} \equiv E_e - E_0$ is the energy splitting between $|0\rangle$ and an excited state $|e\rangle$. $\omega_k$ is the frequency of the $k$-th phonon mode, with the corresponding annihilation (creation) operator $a_k$ ($a_k^\dag$).

For simplicity, let us first consider one ion, say ion $j$, in the chain of $N$ ions. Suppose two beams of laser (with frequencies and wave vectors $\omega_i, \boldsymbol{k}_i, i=1,2$) are shined on the ion to off-resonantly couple states $|0\rangle$ and $|e\rangle$, and $|1\rangle$ and $|e\rangle$, with Rabi frequencies $\Omega_1(t)$ and $\Omega_2(t)$ respectively (see Fig.~\ref{fig:Raman}). This corresponds to the following coupling Hamiltonian
\begin{align}
H' =& \hbar\Omega_1 \cos\left(\boldsymbol{k}_1 \cdot \boldsymbol{r}_j - \omega_1 t - \varphi_1\right) \left(|0\rangle_j\langle e| +|e\rangle_j\langle 0|\right) +\nonumber\\
&\hbar\Omega_2 \cos\left(\boldsymbol{k}_2 \cdot \boldsymbol{r}_j - \omega_2 t - \varphi_2\right) \left(|1\rangle_j\langle e| +|e\rangle_j\langle 1|\right),
\end{align}
where $\Omega_1$ and $\Omega_2$ are chosen to be real. The time dependence of the Rabi frequency has been omitted for convenience. Define $\Delta \equiv \omega_1 - \omega_{0e}$ as the single-photon detuning and $\delta \equiv \omega_1 - \omega_2 -\omega_{01}$ as the two-photon detuning. Here we assume $|\delta| \ll \omega_{01}$ so that we can neglect other two-photon processes between $|0\rangle$ and $|1\rangle$.

Now we perform a unitary transformation characterized by $U=\exp(-iH_0 t/\hbar)$ with
\begin{align}
H_0 =& \hbar\sum_{i\ne j} \left(\omega_{01}|1\rangle_i\langle 1| + \omega_{0e} |e\rangle_i\langle e|\right) + \hbar\sum_k \omega_k a_k^\dag a_k \nonumber\\
&+ \hbar\left(\omega_{01}|1\rangle_j\langle 1| + \omega_1 |e\rangle_j\langle e|\right).
\end{align}
Then the Hamiltonian in the transformed frame, a.k.a. the Hamiltonian in the interaction picture, is given by
\begin{align}
H_I =& U^\dag H U + i \frac{\partial U^\dag}{\partial t}U\nonumber\\
=&-\hbar \Delta |e\rangle_j\langle e| + \frac{\hbar\Omega_1}{2} \Big\{ |e\rangle_j\langle 0|e^{i\left[\boldsymbol{k}_1\cdot \boldsymbol{r}_j(t) -\varphi_1\right]} + h.c. \Big\}\nonumber\\
&  + \frac{\hbar\Omega_2}{2} \Big\{ |e\rangle_j\langle 1|e^{i\left[\boldsymbol{k}_2\cdot \boldsymbol{r}_j(t) -\varphi_2 +\delta \cdot t\right]} + h.c. \Big\},
\end{align}
where $\boldsymbol{r}_j(t)$ is the position operator of ion $j$ at time $t$, under the free evolution of the collective phonon modes. Here we have made the rotating wave approximation (RWA) with the requirement $|\delta|, |\Omega_1|, |\Omega_2| \ll \omega_1, \omega_2$.

\begin{figure}[!tbp]
  \centering
  \includegraphics[width=0.9\linewidth]{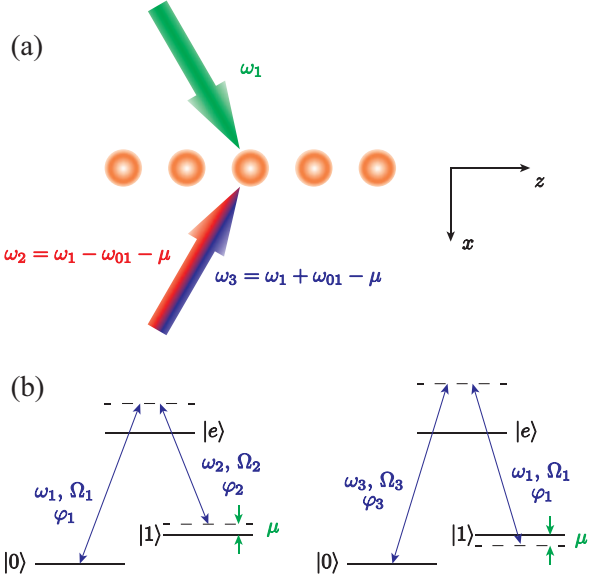}
  \caption{(a) Schematic experimental setup. Three beams are shined on the ion, with one beam in the direction of $\boldsymbol{k}_1$ and the other two red- and blue-detuned beams in the direction of $\boldsymbol{k}_2$. (b) Schematic energy levels and the two pairs of Raman transitions.}\label{fig:three_beam}
\end{figure}

Assume $|\Delta| \gg |\delta|, |\Omega_1|, |\Omega_2|, \gamma_e$ so that the excited state can be adiabatically eliminated, where $\gamma_e$ is the spontaneous emission rate of the excited state. The effective coupling between the state $|0\rangle_j$ and $|1\rangle_j$ is given by
\begin{equation}
H_I^{\mathrm{(eff)}} = \hbar\frac{\Omega_1 \Omega_2}{4\Delta} e^{-i\left[\boldsymbol{\Delta k} \cdot \boldsymbol{r}_j(t) - \delta \cdot t - \Delta \varphi\right]} |0\rangle_j\langle 1| + h.c.\label{eq:H_Raman}
\end{equation}
where $\boldsymbol{\Delta k} \equiv \boldsymbol{k}_1 - \boldsymbol{k}_2$, $\Delta \varphi \equiv \varphi_1 - \varphi_2$. The state $|0\rangle$ and $|1\rangle$ are coupled by an effective Rabi frequency $\Omega_j^{\mathrm{(eff)}} \equiv \Omega_1 \Omega_2/2\Delta$. Later, for simplicity, we drop the superscript and denote the effective Rabi frequency on ion $j$ by $\Omega_j$.

The laser also produces AC Stark shift on the two levels. By suitably choosing the relative intensity of the two laser beams and the detuning $\Delta$ with respect to the excited state, we can make the shifts on the two levels nearly the same \cite{Wineland2003,Campbell2010}. We will discuss more about this effect in Sec.~\ref{sec:result}.

This effective coupling depends on the relative phase of the two laser beams and therefore the fluctuation on their paths. This problem can be solved by adding a third laser beam to form two pairs of Raman transitions, with detuning $\delta = \pm \mu$ and wave vector difference $\pm \Delta k$ along the $x$ direction (see Fig.~\ref{fig:three_beam}). This is known as the phase-insensitive geometry \cite{Lee2005}. We will also briefly discuss the relevance to the phase-sensitive geometry at the end of this subsection.

The effective Rabi frequencies of both pairs are chosen to be $\Omega_j$. Here we assume $\omega_{01}\ll \omega_1, \omega_2, \omega_3$, so that $\Delta k$ is nearly the same for both pairs, with a relative error of the order $\omega_{01}/\omega_1$. Suppose the initial phase differences for the two pairs are $\Delta \varphi_b\equiv\varphi_1-\varphi_2$ and $\Delta \varphi_r\equiv\varphi_3-\varphi_1$. Then the total interaction Hamiltonian can be written as
\begin{widetext}
\begin{align}
H_I^{\mathrm{(eff)}} &= \frac{\hbar\Omega_j}{2} \left[e^{-i\Delta k \cdot x_j(t)} e^{i\mu t}e^{i \Delta \varphi_b} + e^{i\Delta k \cdot x_j(t)} e^{-i\mu t}e^{i \Delta \varphi_r}\right]|0\rangle_j\langle 1| +h.c. \nonumber\\
&= \frac{\hbar\Omega_j}{2} \left[e^{-i\Delta k \cdot x_j(t)} e^{i\mu t}e^{i\varphi_j^{(m)}} e^{i \varphi_j^{(s)}} + e^{i\Delta k \cdot x_j(t)} e^{-i\mu t}e^{-i\varphi_j^{(m)}} e^{i \varphi_j^{(s)}}\right]|0\rangle_j\langle 1| +h.c. \nonumber\\
&= \hbar \Omega_j \cos \left[\mu t + \varphi_j^{(m)} - \Delta k \cdot x_j(t)\right] \left(e^{i \varphi_j^{(s)}} |0\rangle_j\langle 1| + e^{-i \varphi_j^{(s)}} |1\rangle_j\langle 0|\right)\nonumber\\
&= \hbar \Omega_j \cos \left[\mu t + \varphi_j^{(m)} - \Delta k \cdot x_j(t)\right] \left( \sigma_j^x \cos \varphi_j^{(s)} - \sigma_j^y \sin \varphi_j^{(s)}\right),\label{eq:H_twobeam}
\end{align}
\end{widetext}
where $\varphi_j^{(m)} \equiv (\Delta \varphi_b - \Delta \varphi_r)/2$ and $\varphi_j^{(s)} \equiv (\Delta \varphi_b + \Delta \varphi_r)/2$ are called the motional phase and the spin phase \cite{Lee2005}. The subscript $j$ is used to show that these phases pertain to ion $j$. Small fluctuation in beams' paths causes opposite changes in $\Delta \varphi_b$ and $\Delta \varphi_r$, so $\varphi_j^{(s)}$ is robust against fluctuation. On the other hand, $\varphi_j^{(m)}$ does change, but it can be quite stable during one gate time. As we will show later, the gate fidelity is not sensitive to a constant $\varphi_j^{(m)}$ so long as the phase is the same for both ions. Finally we will choose $\varphi_j^{(m)} = 0$ and $\varphi_j^{(s)} = 0$, but for the moment let us keep them in the formulae for completeness.

We further define $\sigma_j^n \equiv \sigma_j^x \cos \varphi_j^{(s)} - \sigma_j^y \sin \varphi_j^{(s)}$ to simplify and drop the superscript on $H_{I}^{(\mathrm{eff})}$:
\begin{equation}
H_I = \hbar \Omega_j(t) \cos \left[\mu t + \varphi_j^{(m)} - \Delta k \cdot x_j(t)\right] \sigma_j^n.
\end{equation}

The transverse position of ion $j$ can be quantized as
\begin{equation}
x_j(t) = \sum_k b_j^k \sqrt{\frac{\hbar}{2 m \omega_k}} \left(a_k e^{-i\omega_k t} + a_k^\dag e^{i \omega_k t}\right),
\end{equation}
where $b_j^k$ ($j=1,2,\cdots,N$) characterizes the $k$-th normalized mode vector of the collective oscillation. The summation over $k$ is limited to the transverse modes along the $x$ direction. This can be done because the small oscillations along $x,y,z$ directions are separable (see Appendix~\ref{app:anharmonic} for more details).

With the Lamb-Dicke parameter $\eta_k \equiv \Delta k \sqrt{\hbar/2m\omega_k}$, we get
\begin{align}
H_I=&\hbar \Omega_j \sigma_j^n \times\nonumber\\
& \cos \Big[\mu t + \varphi_j^{(m)} - \sum_k \eta_k b_j^k \big(a_k e^{-i\omega_k t} + a_k^\dag e^{i\omega_k t}\big)\Big].
\end{align}

We can expand this expression according to the power of $\eta_k$:
\begin{widetext}
\begin{align}
H_I=&\hbar \Omega_j \Bigg[\cos \left(\mu t + \varphi_j^{(m)}\right) + \sin \left(\mu t + \varphi_j^{(m)}\right) \sum_k \eta_k b_j^k \left(a_k e^{-i\omega_k t} + a_k^\dag e^{i\omega_k t}\right)\nonumber\\
& -\frac{1}{2}\cos \left(\mu t + \varphi_j^{(m)}\right) \sum_k \sum_l \eta_k \eta_l b_j^k b_j^l \left(a_k e^{-i\omega_k t} + a_k^\dag e^{i\omega_k t}\right) \left(a_l e^{-i\omega_l t} + a_l^\dag e^{i\omega_l t}\right)\Bigg] \sigma_j^n + O(\eta_k^3).\label{eq:H_expansion}
\end{align}
\end{widetext}

The zeroth order term is a single-qubit operation and commutes with other terms. So we can drop it now and apply a single-qubit rotation after the entangling gate to compensate its effect. Actually for the examples considered in Sec.~\ref{sec:result}, we will show that such a compensation is unnecessary. Here we keep terms up to the second order, but we will show later that the error in the fidelity is of the order $O(\eta_k^4)$.

When the lasers are shined on two ions, we get the interaction-picture Hamiltonian
\begin{align}
H_I =& \sum_{j=j_1,j_2} \sum_k \chi_j(t) \eta_k b_j^k \left(a_k e^{-i\omega_k t} + a_k^\dag e^{i\omega_k t}\right) \sigma_j^n\nonumber\\
& -\frac{1}{2} \sum_{j=j_1,j_2} \sum_k \sum_l \theta_j(t) \eta_k \eta_l b_j^k b_j^l \times \nonumber\\
& \left(a_k e^{-i\omega_k t} + a_k^\dag e^{i\omega_k t}\right)\left(a_l e^{-i\omega_l t} + a_l^\dag e^{i\omega_l t}\right) \sigma_j^n,\label{eq:H_expansion_2}
\end{align}
where the summation of $j$ is over the two ions and
\begin{equation}
\chi_j(t) \equiv \hbar \Omega_j \sin\left(\mu t + \varphi_j^{(m)}\right),
\end{equation}
\begin{equation}
\theta_j(t) \equiv \hbar \Omega_j \cos\left(\mu t + \varphi_j^{(m)}\right).
\end{equation}

Unitary evolution in the interaction picture is obtained by the Magnus expansion
\begin{align}
U_I(\tau) \approx& \exp\Bigg(i \sum_j \left[\phi_j(\tau) + \psi_j(\tau)\right]\sigma_j^n\nonumber\\
& \qquad\qquad\qquad+ i \sum_{i<j}\Theta_{ij}(\tau)\sigma_i^n \sigma_j^n\Bigg),\label{eq:U_expansion}
\end{align}
where
\begin{equation}
\phi_j(\tau) = -i \sum_k \left[\alpha_j^k(\tau)a_k^\dag - {\alpha_j^k}^* (\tau) a_k\right],
\end{equation}
\begin{equation}
\label{eq:alpha}
\alpha_j^k(\tau) = -\frac{i}{\hbar} \eta_k b_j^k \int_0^\tau \chi_j(t) e^{i\omega_k t} dt,
\end{equation}
\begin{equation}
\label{eq:psi}
\psi_j(\tau) = \sum_k \lambda_j^k(\tau) \left(a_k^\dag a_k + \frac{1}{2}\right),
\end{equation}
\begin{equation}
\label{eq:lambda}
\lambda_j^k(\tau) = \frac{1}{\hbar}\left(\eta_k b_j^k\right)^2 \int_0^{\tau} \theta_j(t)dt,
\end{equation}
describe the coupling between the spin and phonon modes, and
\begin{widetext}
\begin{equation}
\label{eq:phi}
\Theta_{ij}(\tau)=\frac{1}{\hbar^2}\sum_k \eta_k^2 b_i^k b_j^k \int_0^\tau dt_1 \int_0^{t_1}dt_2 \left[\chi_i(t_1) \chi_j(t_2) + \chi_j(t_1) \chi_i(t_2)\right]\sin \left[\omega_k (t_1-t_2)\right].
\end{equation}
\end{widetext}
is the coupling between the two spins. Roughly speaking, the $\phi_j$ terms are displacement operations on the phonon modes conditioned on the spin state of each ion, and the $\psi_j$ terms are single-spin rotations conditioned on the phonon numbers of each mode. We need to suppress these terms while maintain a large spin-spin coupling to realize the entangling gate. Here again we keep terms up to the second order in $\eta_k$ and retain only diagonal terms in $\psi_j(\tau)$ [Eq.~(\ref{eq:psi})]. An error analysis is performed in Sec.~\ref{sec:error} and Appendix~\ref{app:higher_order}.  In the above derivation we have also dropped a global phase, which has no effect on the entangling gate.

If the effective Rabi frequencies of the laser beams on the two ions are always proportional, e.g. when the lasers come from a single beam through a beam splitter, the expression of $\Theta_{ij}$ can be simplified as
\begin{align}
\label{eq:phi_simple}
\Theta_{ij}(\tau)=&\frac{2}{\hbar^2}\sum_k \eta_k^2 b_i^k b_j^k \int_0^\tau dt_1 \times\nonumber\\
&\int_0^{t_1}dt_2 \chi_i(t_1) \chi_j(t_2)\sin \left[\omega_k (t_1-t_2)\right].
\end{align}
In this way we recover Eq.~(2) of Ref.~\cite{Zhu2006}.

In the above derivation we assumed a phase-insensitive laser configuration. It is also possible to use the phase-sensitive geometry for the entangling gate with the spin phase being cancelled by a Ramsey-like gate design \cite{Lee2005,Gaebler2016}. However, due to the difference in the resulting Hamiltonian and hence the different commutation relation, a similar expansion in the Lamb-Dicke parameter leads to infinitely more terms. It seems to us that there is no easy justification to throw away these terms for the model we are considering, so we will not go into further details here. Nevertheless, except for the higher order terms in Lamb-Dicke parameters, our other analyses in Sec.~\ref{sec:result} can still be applied to the phase-sensitive setup.

\subsection{XX Entangling Gate and Fidelity}
If $\alpha^k_j(\tau)=0$ and $\lambda_j^k(\tau)=0$ for all the modes and both of the ions, $\varphi_j^{(s)}=0$ for both ions, and $\Theta_{ij} = \pi/4$ for the ions of interest, $i$ and $j$, the time evolution operator will be an ideal XX entangling gate. In the basis of $|+\rangle_i|+\rangle_j$, $|+\rangle_i|-\rangle_j$, $|-\rangle_i|+\rangle_j$, $|-\rangle_i|-\rangle_j$ where $|\pm\rangle=\frac{1}{\sqrt{2}}(|0\rangle\pm|1\rangle)$, we have
\begin{equation}
U_{\mathrm{ideal}} = e^{i\pi\sigma_i^x \sigma_j^x/4} = \left(
\begin{array}{cccc}
e^{i\pi/4} & 0 & 0 & 0\\
0 & e^{-i\pi/4} & 0 & 0\\
0 & 0 & e^{-i\pi/4} & 0\\
0 & 0 & 0 & e^{i\pi/4}
\end{array}\right).
\end{equation}
The subscript $ij$ and the dependence on $\tau$ have been dropped.

If the initial internal state is $|\Psi_0\rangle$ and the vibrational modes are in the thermal state $\rho_{\mathrm{th}}$ with a temperature $T$, the ideal final state is $U_{\mathrm{ideal}}|\Psi_0\rangle$, while the actual state we get is $\rho=\mathrm{tr}_m [U|\Psi_0\rangle\langle \Psi_0|\otimes \rho_{\mathrm{th}} U^\dag]$, where $\mathrm{tr}_m$ means the partial trace over all the motional modes. Then we can use the fidelity $F =\langle\Psi_0|U_{\mathrm{ideal}}^\dag \rho U_{\mathrm{ideal}}|\Psi_0\rangle$ to characterize the similarity between these two states and therefore between the ideal and the real gates.

However, the above method depends on the initial state $|\Psi_0\rangle$. For a state-independent measure of the similarity between $U$ and $U_{\mathrm{ideal}}$, we can use the average gate fidelity \cite{Nielsen2002}
\begin{equation}
\overline{F} =\int d\Psi\langle\Psi|U_{\mathrm{ideal}}^\dag \mathrm{tr}_m [U|\Psi\rangle\langle \Psi|\otimes \rho_{\text{th}} U^\dag] U_{\mathrm{ideal}}|\Psi\rangle,
\end{equation}
where the integration is over the Fubini-Study measure \cite{bengtsson_zyczkowski_2006}. For the moment we assume the spin phases $\varphi_j^{(s)}=0$ for the two ions, i.e. $\hat{n}=\hat{x}$, $\sigma_j^n = \sigma_j^x$. Later we will discuss the effects of nonzero spin phases in Sec.~\ref{sec:parameters}.

Let us express $U$ in the above basis:
\begin{equation}
U = \left(
\begin{array}{cccc}
e^{i\Phi_{00}} & 0 & 0 & 0\\
0 & e^{i\Phi_{01}} & 0 & 0\\
0 & 0 & e^{i\Phi_{10}} & 0\\
0 & 0 & 0 & e^{i\Phi_{11}}
\end{array}\right),
\end{equation}
where $\Phi_{00}=\phi_i + \psi_i + \phi_j + \psi_j +\Theta_{ij}$, $\Phi_{01}=\phi_i + \psi_i - \phi_j - \psi_j - \Theta_{ij}$, $\Phi_{10}=- \phi_i -\psi_i+ \phi_j +\psi_j-\Theta_{ij}$, $\Phi_{11}=-\phi_i -\psi_i - \phi_j -\psi_j +\Theta_{ij}$ are the phases gained by the $|+\rangle_i|+\rangle_j$, $|+\rangle_i|-\rangle_j$, $|-\rangle_i|+\rangle_j$, $|-\rangle_i|-\rangle_j$ states, respectively. Note that they are actually operators in the subspace of phonon modes.

Accurate up to second order diagonal terms in $\eta_k$, we have
\begin{align}
e^{i\Phi_{00}} \approx& e^{i\Theta_{ij}}\prod_k D_k \left(\alpha_i^k(\tau)+\alpha_j^k(\tau)\right) \times\nonumber\\
&\left\{1+ i \sum_l \left[\lambda_i^l(\tau) + \lambda_j^l(\tau)\right] \left(a_l^\dag a_l + \frac{1}{2}\right)\right\},\label{eq:Phi00}
\end{align}
\begin{align}
e^{i\Phi_{01}} \approx& e^{-i\Theta_{ij}}\prod_k D_k \left(\alpha_i^k(\tau)-\alpha_j^k(\tau)\right) \times\nonumber\\
&\left\{1+ i \sum_l \left[\lambda_i^l(\tau) - \lambda_j^l(\tau)\right] \left(a_l^\dag a_l + \frac{1}{2}\right)\right\},\label{eq:Phi01}
\end{align}
\begin{align}
e^{i\Phi_{10}} \approx& e^{-i\Theta_{ij}}\prod_k D_k \left(-\alpha_i^k(\tau)+\alpha_j^k(\tau)\right) \times\nonumber\\
&\left\{1- i \sum_l \left[\lambda_i^l(\tau) - \lambda_j^l(\tau)\right] \left(a_l^\dag a_l + \frac{1}{2}\right)\right\},\label{eq:Phi10}
\end{align}
\begin{align}
e^{i\Phi_{11}} \approx& e^{i\Theta_{ij}}\prod_k D_k \left(-\alpha_i^k(\tau)-\alpha_j^k(\tau)\right) \times\nonumber\\
&\left\{1- i \sum_l \left[\lambda_i^l(\tau) + \lambda_j^l(\tau)\right] \left(a_l^\dag a_l + \frac{1}{2}\right)\right\},\label{eq:Phi11}
\end{align}
where $D_k \left(\alpha\right) \equiv \exp(\alpha a_k^\dag -\alpha^* a_k)$ is the displacement operator of the $k$-th mode.

For an arbitrary operator $\rho_0$ (not necessarily Hermitian)
\begin{equation}
\rho_0 = \left(
\begin{array}{cccc}
\rho_{00,00} & \rho_{00,01} & \rho_{00,10} & \rho_{00,11}\\
\rho_{01,00} & \rho_{01,01} & \rho_{01,10} & \rho_{01,11}\\
\rho_{10,00} & \rho_{10,01} & \rho_{10,10} & \rho_{10,11}\\
\rho_{11,00} & \rho_{11,01} & \rho_{11,10} & \rho_{11,11}\\
\end{array}
\right),
\end{equation}
lengthy but straightforward calculation shows that
\begin{widetext}
\begin{align}
\rho &= \mathrm{tr}_m [U\rho_0 \otimes \rho_{\mathrm{th}} U^\dag] \nonumber\\
&\approx \left(
\begin{array}{cccc}
\rho_{00,00} & \Gamma_j \Lambda_j e^{2i\Theta_{ij}-i\epsilon}\rho_{00,01} & \Gamma_i \Lambda_i e^{2i\Theta_{ij}+i\epsilon}\rho_{00,10} & \Gamma_+ \Lambda_+ \rho_{00,11}\\
\Gamma_j \Lambda_j^* e^{-2i\Theta_{ij}+i\epsilon}\rho_{01,00} & \rho_{01,01} & \Gamma_- \Lambda_- \rho_{01,10} & \Gamma_i \Lambda_i e^{-2i\Theta_{ij}-i\epsilon} \rho_{01,11}\\
\Gamma_i \Lambda_i^* e^{-2i\Theta_{ij}-i\epsilon} \rho_{10,00} & \Gamma_- \Lambda_-^* \rho_{10,01} & \rho_{10,10} & \Gamma_j \Lambda_j e^{-2i\Theta_{ij}+i\epsilon} \rho_{10,11}\\
\Gamma_+ \Lambda_+^* \rho_{11,00} & \Gamma_i \Lambda_i^* e^{2i\Theta_{ij}+i\epsilon} \rho_{11,01} & \Gamma_j \Lambda_j^* e^{2i\Theta_{ij}-i\epsilon} \rho_{11,10} & \rho_{11,11}
\end{array}
\right),\label{eq:rho}
\end{align}
\end{widetext}
where $\epsilon = 2\sum_k\mathrm{Im}(\alpha_i^k {\alpha_j^k}^*)$,
\begin{equation}
\Gamma_{i(j)} = \exp\left[-2\sum_k \left|\alpha_{i(j)}^k\right|^2 \coth\left(\frac{\hbar \omega_k}{2k_B T}\right)\right],
\end{equation}
\begin{equation}
\Gamma_{\pm} = \exp\left[-2\sum_k \left|\alpha_i^k\pm\alpha_j^k\right|^2 \coth\left(\frac{\hbar \omega_k}{2k_B T}\right)\right],
\end{equation}
\begin{equation}
\Lambda_{i(j)} = 1 + i \sum_k \lambda_{i(j)}^k\coth \frac{\hbar \omega_k}{2k_B T},
\end{equation}
\begin{equation}
\Lambda_{\pm} = 1 + i \sum_k \left(\lambda_i^k \pm \lambda_j^k\right)\coth \frac{\hbar \omega_k}{2k_B T}.
\end{equation}
We have used the following formulae in the derivation:
\begin{equation}
D(\alpha)D(\beta) = e^{(\alpha\beta^*-\alpha^*\beta)/2}D(\alpha+\beta),
\end{equation}
\begin{equation}
\mathrm{tr}\left[D(\alpha)\rho_{\mathrm{th}}\right]=\exp\left[-\frac{|\alpha|^2}{2}\coth\left(\frac{\hbar\omega}{2k_B T}\right)\right],
\end{equation}
\begin{equation}
\mathrm{tr}\left[\left(a^\dag a + \frac{1}{2}\right)\rho_{\mathrm{th}}\right] = \frac{1}{2}\coth\left(\frac{\hbar\omega}{2k_B T}\right).
\end{equation}

The average gate fidelity can then be written as \cite{Nielsen2002}
\begin{equation}
\label{eq:fidelity_average}
\overline{F} = \frac{\sum_l \mathrm{tr} \left\{U_{\mathrm{ideal}} W_l^\dag U_{\mathrm{ideal}}^\dag \mathrm{tr}_m \left[U W_l \otimes \rho_{\mathrm{th}} U^\dag \right]\right\}+d^2}{d^2(d+1)},
\end{equation}
where $d=4$ and $\{W_l\}$ is an orthogonal basis of $4\times 4$ unitary operators such that $\mathrm{tr} [W_k^\dag W_l] = \delta_{kl}d$. Here we can choose $\{W_l\} = \{V_1 \otimes V_2 | V_1, V_2 = I,\sigma_x,\sigma_y,\sigma_z\}$. Using Eq.~(\ref{eq:rho}) we finally obtain
\begin{align}
\label{eq:fidelity_general}
\overline{F} \approx& \frac{1}{10}\big[4 + 2\Gamma_i \sin(2\Theta_{ij}+\epsilon) + 2\Gamma_j \sin(2\Theta_{ij}-\epsilon) \nonumber\\
&\qquad\qquad+ \Gamma_+ + \Gamma_-\big].
\end{align}

$\lambda_j^k$ terms [Eq.~(\ref{eq:lambda})] appear quadratically in the fidelity, hence its contribution is $O(\eta_k^4)$ and is neglected. If $\Omega \lesssim \mu, \omega_k$ and the average phonon number for a typical mode is $\bar{n}$, the error from neglecting higher order terms is of the order $\eta_k^4 (2\bar{n}+1)^2$. The fact that there are $N$ independent transverse modes has already been included because the coefficient for each mode is also modulated by the $b_j^k$ vectors, which are normalized to 1.

Suppose the laser intensities on the two ions are always proportional and that their phases are locked such that $\varphi_i^{(m)}=\varphi_j^{(m)}=0$, then we get $\epsilon=0$. The above expression can be simplified as
\begin{equation}
\label{eq:fidelity}
\overline{F} = \frac{1}{10}\left[4 + 2(\Gamma_i + \Gamma_j) \sin 2\Theta_{ij} + \Gamma_+ + \Gamma_-\right].
\end{equation}
This average gate fidelity is slightly higher than Eq.~(3) of Ref.~\cite{Zhu2006}, where a special initial state is used.

Also notice that if $\Theta_{ij}=-\pi/4$, the gate is close to another ideal entangling gate $\exp(-i\pi\sigma_i^x \sigma_j^x/4)$, which is different from $U_{\mathrm{ideal}}$ only by local operations. In this case the gate fidelity can be calculated in a similar way and the final result is
\begin{equation}
\label{eq:fidelity_neg}
\overline{F} = \frac{1}{10}\left[4 - 2(\Gamma_i + \Gamma_j) \sin 2\Theta_{ij} + \Gamma_+ + \Gamma_-\right].
\end{equation}
From now on, by fidelity we mean the average gate fidelity if not specifically mentioned. We will drop the overline on $F$ for convenience.

In the experiment, we can set the laser beams on the two ions to be the same. We can divide the laser sequence into $n_{\textrm{seg}}$ equal segments and in each segment let the Rabi frequency be a constant. Define a real column vector $\boldsymbol{\Omega} = (\Omega_1, \Omega_2, \cdots, \Omega_{n_{\textrm{seg}}})^T$ corresponding to the Rabi frequency of each segment, and we get
\begin{align}
\alpha_j^k(\tau) = \boldsymbol{A}_j^k \boldsymbol{\Omega}, \quad
\Theta_{ij} = \boldsymbol{\Omega}^T \boldsymbol{\gamma}' \boldsymbol{\Omega},
\end{align}
where $\boldsymbol{A}_j^k$ is a row vector whose $n$-th component is
\begin{equation}
\boldsymbol{A}_j^k (n) = -i \eta_k b_j^k \int_{(n-1)\tau/n_{\textrm{seg}}}^{n\tau/n_{\textrm{seg}}} \sin \mu t \cdot e^{i \omega_k t} dt,
\end{equation}
and $\boldsymbol{\gamma}'$ is an $n_{\textrm{seg}}$ by $n_{\textrm{seg}}$ matrix whose $(p,q)$ component is
\begin{widetext}
\begin{equation}
\boldsymbol{\gamma}'(p,q) = \left\{
\begin{array}{ll}
\displaystyle 2\sum_k \eta_k^2 b_i^k b_j^k \int_{(p-1)\tau/n_{\textrm{seg}}}^{p\tau/n_{\textrm{seg}}} dt_1 \int_{(q-1)\tau/n_{\textrm{seg}}}^{q\tau/n_{\textrm{seg}}} dt_2 \sin \mu t_1 \sin \mu t_2 \sin [\omega_k (t_1 - t_2)] & (p>q)\\
\displaystyle  2\sum_k \eta_k^2 b_i^k b_j^k \int_{(p-1)\tau/n_{\textrm{seg}}}^{p\tau/n_{\textrm{seg}}} dt_1 \int_{(p-1)\tau/n_{\textrm{seg}}}^{t_1} dt_2 \sin \mu t_1 \sin \mu t_2 \sin [\omega_k (t_1 - t_2)] & (p=q)\\
0 & (p<q)
\end{array}\right..
\end{equation}
\end{widetext}
We can further define a symmetric matrix $\boldsymbol{\gamma} \equiv (\boldsymbol{\gamma}' + \boldsymbol{\gamma}'^T)/2$ such that $\Theta_{ij} = \boldsymbol{\Omega}^T \boldsymbol{\gamma}' \boldsymbol{\Omega} = \boldsymbol{\Omega}^T \boldsymbol{\gamma} \boldsymbol{\Omega}$. By suitably scaling $\boldsymbol{\Omega}$, we can always set $\Theta_{ij} = \pm \pi/4$. Then in the limit of small $\alpha$ (high fidelity), the fidelity can be approximated as
\begin{align}
F &\approx 1 - \frac{4}{5}\sum_k \left(|\alpha_i^k|^2 + |\alpha_j^k|^2\right) \coth \frac{\hbar \omega_k}{2k_B T}\nonumber\\
&= 1-\frac{4}{5}\boldsymbol{\Omega}^T \left[\sum_k \left( {\boldsymbol{A}_i^k}^\dag \boldsymbol{A}_i^k + {\boldsymbol{A}_j^k}^\dag \boldsymbol{A}_j^k \right)\coth \frac{\hbar \omega_k}{2k_B T}\right]\boldsymbol{\Omega}\nonumber\\
&\equiv 1- \frac{4}{5}\boldsymbol{\Omega}^T \boldsymbol{M} \boldsymbol{\Omega}.\label{eq:fidelity_approx}
\end{align}

By definition, $\boldsymbol{M}$ is a Hermitian matrix, but actually we can express it in a real symmetric form:
\begin{align}
\boldsymbol{\Omega}^T \boldsymbol{M} \boldsymbol{\Omega} &= \frac{1}{2}\left(\boldsymbol{\Omega}^T \boldsymbol{M} \boldsymbol{\Omega} + \boldsymbol{\Omega}^T \boldsymbol{M}^T \boldsymbol{\Omega}\right) \nonumber\\
&= \frac{1}{2}\left(\boldsymbol{\Omega}^T \boldsymbol{M} \boldsymbol{\Omega} + \boldsymbol{\Omega}^T \boldsymbol{M}^* \boldsymbol{\Omega}\right) \nonumber\\
&= \boldsymbol{\Omega}^T \mathrm{Re}[\boldsymbol{M}] \boldsymbol{\Omega}.
\end{align}

Now we want to minimize $\boldsymbol{\Omega}^T \boldsymbol{M} \boldsymbol{\Omega}$ under the constraint $\boldsymbol{\Omega}^T \boldsymbol{\gamma} \boldsymbol{\Omega} = \pm \pi/4$. For this purpose, we use the method of Lagrange multiplier and consider the optimization of $f(\boldsymbol{\Omega},\lambda)=\boldsymbol{\Omega}^T \boldsymbol{M} \boldsymbol{\Omega}-\lambda(\boldsymbol{\Omega}^T \boldsymbol{\gamma} \boldsymbol{\Omega} \mp \pi/4)$:
\begin{equation}
\left\{
\begin{array}{l}
\boldsymbol{M}\boldsymbol{\Omega} - \lambda \boldsymbol{\gamma} \boldsymbol{\Omega} = 0\\
\boldsymbol{\Omega}^T \boldsymbol{\gamma} \boldsymbol{\Omega} = \pm \pi/4
\end{array}\right..\label{eq:optimization}
\end{equation}
We only need to solve this generalized eigenvalue problem and find the eigenvalue with the smallest absolute value. The corresponding eigenvector, with suitable normalization, gives us the optimal $\boldsymbol{\Omega}$. (See also Appendix A of Ref.~\cite{lin2010quantum}.)

We remark that for realistic experimental parameters, the effective Rabi frequency cannot be too large. This means that the above optimization should be performed under another inequality constraint. This problem is generally hard to solve, so instead we use the method mentioned above and then discard solutions with unrealistic $|\boldsymbol{\Omega}|$.
\section{Systematic Errors and Experimental Noise}
\label{sec:result}
The gate fidelity realized in the experiment is always less than 1. This is due to the approximations in the formulation, imperfections in the gate design, as well as noise and errors in the experiment. In this section we analyze these sources of errors in detail.

\begin{figure}[!tbp]
  \centering
  \includegraphics[width=0.6\linewidth]{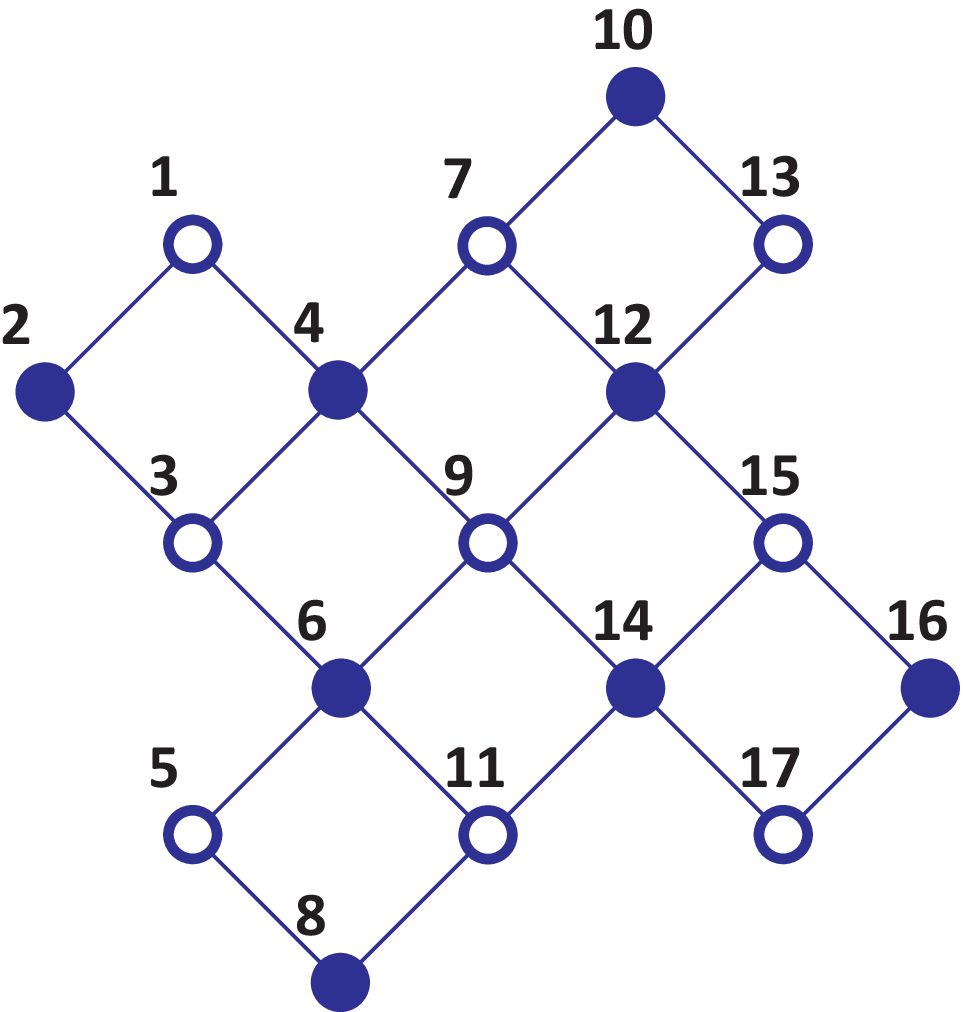}\\
  \caption{A 17-qubit surface code layout. The open circles represent the data qubits and the filled circles represent the syndrome qubits. Labels 1-17 corresponds to the real order of qubits in the 1D chain.}\label{fig:surface_code}
\end{figure}

In order to estimate the influence of each error term, we consider a specific example of mapping a 17-qubit surface code for quantum error correction into a linear chain of ${}^{171}\mathrm{Yb}^+$ ions \cite{Tomita2014,1710.01378} (see Fig.~\ref{fig:surface_code} for the mapping).
For this purpose, diamond norm may be a better measure of the gate performance, but we focus on average gate fidelity here as it is easier to treat theoretically. We will discuss their difference in Sec.~\ref{sec:dnorm}. For realistic parameters, we choose $\omega_x=\omega_y=2\pi\times 3\,$MHz, and consider a chain of 19 ions with the two ions at the ends only used for cooling. An anharmonic potential is applied along the $z$ axis, which is specified by $l_0=40\,\mu$m and $\gamma_4=4.3$ (see Appendix~\ref{app:anharmonic} for the definition). In this way the central 17 ions will have a nearly uniform spacing with an average of $d_{\mathrm{av}} = 8.3\,\mu$m and a relative standard deviation of 2.3\%.

Under these conditions, the spectrum of the transverse normal modes is very narrow (within 0.9\% of $\omega_x$). Hence it is possible to use sideband cooling method to cool the transverse motion down to about 0.5 phonon per mode or less. Doppler cooling can also be used if the trapping can be stronger. For counter-propagating laser beams along the $\pm x$ directions with $\lambda=355\,$nm \cite{Campbell2010}, we have a detuning $\Delta\approx 2\pi\times 33\,$THz and $\Delta k=2k$. (Actually there are two excited states with a fine-structure splitting of $2\pi\times 100\,$THz, and the laser detuning is specially chosen to minimize the differential AC Stark shift. We will come back to this point when discussing about the AC Stark shift; but otherwise we will just use one value of $\Delta$ to estimate the order of magnitude for the other error terms.) The Lamb-Dicke parameter is then $\eta_k\approx 0.11$ for all the transverse modes.

\subsection{Optimized Gate Design and Sensitivity to Tunable Parameters}
\label{sec:parameters}
In order to perform the stabilizer measurement in the surface code, we need to achieve two-qubit entangling gates between nearest neighbor qubits in Fig.~\ref{fig:surface_code}, that is, ion pairs with one, three and five ion separations in the linear chain. To find the optimal parameters for a high-fidelity gate, we use Eqs.~(\ref{eq:fidelity_approx}) and (\ref{eq:optimization}) to estimate the gate fidelity and to solve the optimal pulse sequence. We then scan the gate time $\tau$, detuning $\mu$ and number of segments $n_{\mathrm{seg}}$ to find a combination with the desired fidelity.

\begin{figure}[!tbp]
  \centering
  \includegraphics[width=\linewidth]{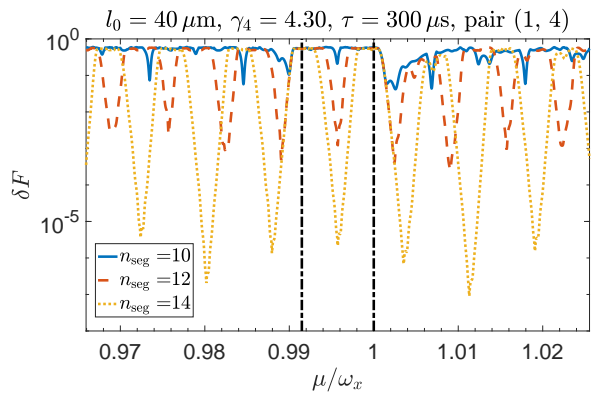}\\
  \caption{Infidelity for an entangling gate between ion 1 and ion 4 as a function of the detuning $\mu$. Here gate time $\tau=300\,\mu$s and 3 segment numbers $n_{\mathrm{seg}}=10,12,14$ are used. The vertical dash-dot lines give the range of the spectrum of the transverse normal modes.}\label{fig:fidelity_scan}
\end{figure}

For example, Fig.~\ref{fig:fidelity_scan} shows the gate infidelity ($\delta F\equiv 1-F$) for the entangling gate between ion 1 and ion 4 as a function of detuning $\mu$ for a fixed gate time $\tau=300\,\mu$s and three possible segment numbers $n_{\mathrm{seg}}=10, 12, 14$.
As we can see, increasing the number of segments generally reduces the gate infidelity. We also notice that there are multiple local minima in the gate infidelity. Therefore, we do not attempt to find the ``best'' solution, but rather look for solutions that are ``good enough''. That is, the solution needs to achieve high gate fidelity in the ideal case, and it should also be robust against errors in these control parameters, which may arise from imperfect calibration, finite resolution or random fluctuation in the experiment. Specifically, we perturb the gate parameters at local minima of plots similar to Fig.~\ref{fig:fidelity_scan} when scanning these parameters and keep the ones that are most insensitive to the noise. We will assume that these noises are ``slow'' such that they stay constant during one gate period. This assumption is reasonable because typically the high-frequency noise will be weak in the experiment. For instance, Ref.~\cite{Ballance2017} considers the influence of high-frequency noise in a two-ion crystal and the experimental noise level is found to be about one order of magnitude lower than what is allowed for an error of $10^{-4}$. Also note that the same technique to optimize the gate design has been applied in Ref.~\cite{1710.01378}, but the number of segments and the gate time we use here are generally larger because of this additional requirement of robustness.

Below we show the results for ion pairs with three typical separations. For experimentally achievable effective Rabi frequencies, we only present solutions satisfying $|\Omega(t)|<2\pi\times 1\,$MHz at all times.
\begin{itemize}
\item Ion 5 and ion 6 (separation 1):

We use $n_{\mathrm{seg}}=10$ segments and $\tau = 80.4\,\mu$s. Laser sequence $\boldsymbol{\Omega}_0$ is optimized for $\mu_0=0.995\omega_x$ (Fig.~\ref{fig:Omega_5_6}). For the sensitivity to control parameters, in Fig.~\ref{fig:ion_5_6} we show how the gate infidelity changes under a shift in detuning $\mu$ of $2\pi\times 1\,$kHz, in the global laser intensity $\boldsymbol{\Omega}$ of $1\%$, in gate time $\tau$ of $0.4\,\mu$s, as well as the effect of a nonzero $\varphi^{(m)}$. (See Eq.~(\ref{eq:H_twobeam}) for the definition. Here the motional phase is assumed to be equal for both ions.) For parameters fluctuating inside these ranges, the gate infidelity is always below $10^{-3}$.
    \begin{figure}[!tbp]
        \centering
        \includegraphics[width=0.9\linewidth]{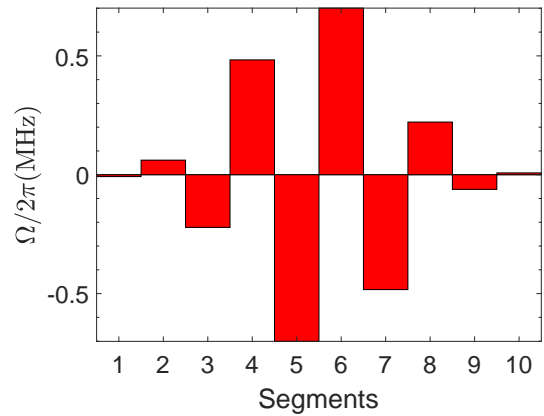}
        \caption{Optimized effective Rabi frequency sequence $\boldsymbol{\Omega}_0$ on ion 5 and ion 6 for $n_{\mathrm{seg}}=10$, detuning $\mu_0=0.995\omega_x$ and gate time $\tau=80.4\,\mu$s. Here we allow the Rabi frequency to take negative values by adding a phase shift of $\pi$. If such a phase shift is not available, we can look for other solutions where all the Rabi frequencies are positive. Some examples are shown in Ref.~\cite{1710.01378}.}\label{fig:Omega_5_6}
    \end{figure}
    \begin{figure}[!tbp]
      \centering
      \includegraphics[width=0.9\linewidth]{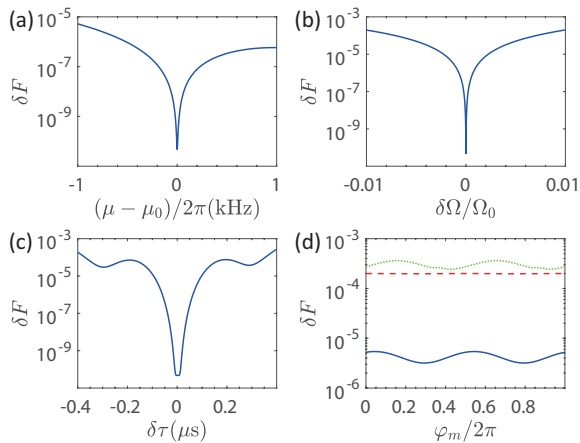}
      \caption{Parameter sensitivity for the entangling gate between ion 5 and ion 6. (a) Infidelity as a function of shift in detuning. (b) Infidelity as a function of relative shift in laser intensity. Here we assume that the frequency of the noise is low so that the laser intensities of all the segments are shifted by the same percentage. (c) Infidelity as a function of shift in gate time $\tau$. (d) Dependence on the motional phase $\varphi^{(m)}$. Here we consider $\varphi_i^{(m)}=\varphi_j^{(m)}$ between 0 and $2\pi$. Solid blue, dashed red and dotted green curves are the maximal infidelity for a shift of $1\,$kHz in detuning $\mu$, a 1\% change in Rabi frequency, and $0.4\,\mu$s change in total gate time, respectively.}\label{fig:ion_5_6}
    \end{figure}

\item Ion 1 and ion 4 (separation 3):

We use $n_{\mathrm{seg}}=17$ segments and $\tau = 250\,\mu$s. Laser sequence $\boldsymbol{\Omega}_0$ is optimized for $\mu_0=0.997\omega_x$ (Fig.~\ref{fig:Omega_1_4}), but then for the robustness under fluctuation in detuning (where positive and negative shifts have asymmetric effect), the gate is performed at the detuning $\mu_0'=\mu_0+2\pi\times 0.8\,$kHz with a slight rescaling of the laser intensity. (See Appendix~\ref{app:accumulation} for more details about this rescaling, which aims to reduce the accumulation of errors when multiple gates are applied.) Therefore in Fig.~\ref{fig:ion_1_4} the smallest infidelity does not always appear at the center of the parameter range.
    \begin{figure}[!tbp]
      \centering
      \includegraphics[width=0.9\linewidth]{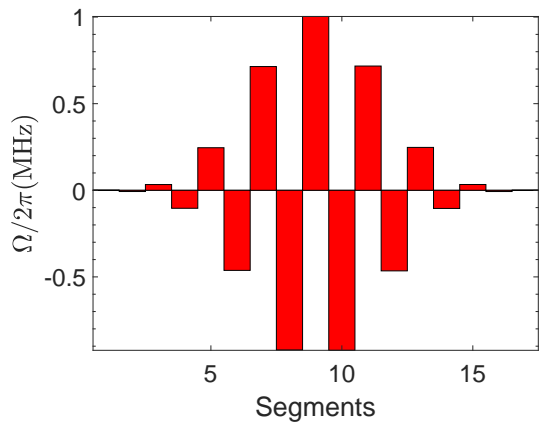}
      \caption{Optimized laser sequence $\boldsymbol{\Omega}_0$ on ion 1 and ion 4 for $n_{\mathrm{seg}}=17$, detuning $\mu_0=0.997\omega_x$ and gate time $\tau=250\,\mu$s.}\label{fig:Omega_1_4}
    \end{figure}
    \begin{figure}[!tbp]
      \centering
      \includegraphics[width=0.9\linewidth]{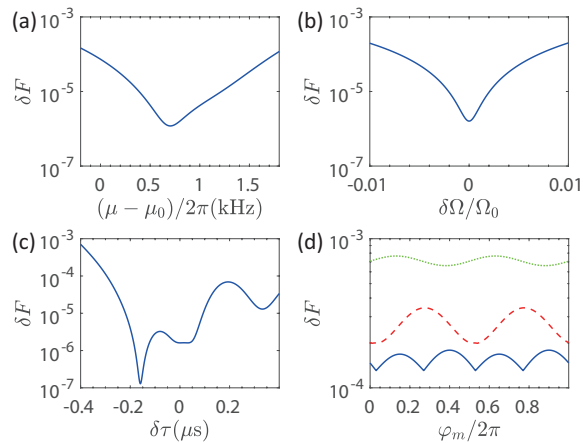}
      \caption{Parameter sensitivity for the entangling gate between ion 1 and ion 4. (a) Infidelity as a function of shift in detuning. (b) Infidelity as a function of relative shift in Rabi frequency. We assume the laser intensities of all the segments are shifted by the same percentage. (c) Infidelity as a function of shift in gate time $\tau$. (d) Consider $\varphi_i^{(m)}=\varphi_j^{(m)}$ between 0 and $2\pi$. Solid blue, dashed red and dotted green curves are the maximal infidelity for a shift of $1\,$kHz in detuning $\mu$, a 1\% change in intensity, and $0.4\,\mu$s change in total gate time, respectively.}\label{fig:ion_1_4}
    \end{figure}

\item Ion 9 and ion 14 (separation 5):

We use $n_{\mathrm{seg}}=24$ segments and $\tau = 482\,\mu$s. Laser sequence $\boldsymbol{\Omega}_0$ is optimized for $\mu_0=0.997\omega_x$ (Fig.~\ref{fig:Omega_9_14}). For the robustness under fluctuation in parameters, we then work at the detuning $\mu_0'=\mu_0-2\pi\times 0.5\,$kHz. Gate infidelity under shifts in parameters are shown in Fig.~\ref{fig:ion_9_14}.
    \begin{figure}[!tbp]
      \centering
      \includegraphics[width=0.9\linewidth]{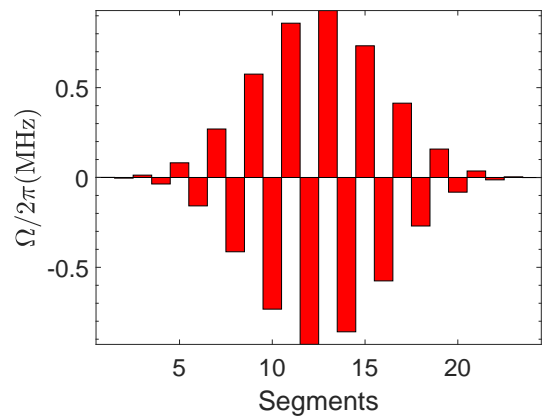}
      \caption{Optimized laser sequence $\boldsymbol{\Omega}_0$ on ion 9 and ion 14 for $n_{\mathrm{seg}}=24$, detuning $\mu_0=0.997\omega_x$ and gate time $\tau=482\,\mu$s.}\label{fig:Omega_9_14}
    \end{figure}
    \begin{figure}[!tbp]
      \centering
      \includegraphics[width=0.9\linewidth]{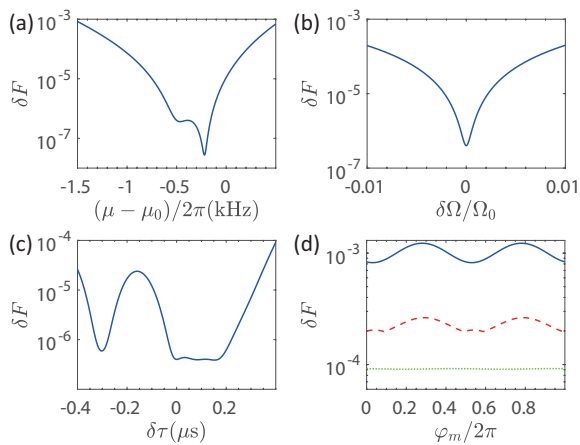}
      \caption{Parameter sensitivity for the entangling gate between ion 9 and ion 14. (a) Infidelity as a function of shift in detuning. (b) Infidelity as a function of relative shift in Rabi frequency. (c) Infidelity as a function of shift in gate time $\tau$. (d) Consider $\varphi_i^{(m)}=\varphi_j^{(m)}$ between 0 and $2\pi$. Solid blue, dashed red and dotted green curves are the maximal infidelity for a shift of $1\,$kHz in detuning $\mu$, a 1\% change in intensity, and $0.4\,\mu$s change in total gate time, respectively.}\label{fig:ion_9_14}
    \end{figure}
\end{itemize}

As we can see in Figs.~\ref{fig:ion_5_6}, \ref{fig:ion_1_4} and \ref{fig:ion_9_14}, a nonzero but constant $\varphi_i^{(m)} = \varphi_j^{(m)}$ does not influence the fidelity significantly. This justifies the use of the phase-insensitive setup, which suppresses the fluctuation in $\varphi^{(s)}$ but allows $\varphi^{(m)}$ to change over different gates. Nevertheless, we still need to set $\varphi^{(s)}=0$ initially for the desired XX entangling gate: by taking $U=\exp(i\pi\sigma_i^{n}\sigma_j^{n}/4)$ in Eq.~(\ref{eq:fidelity_average}) with small spin phases $\varphi_i^{(s)}$ and $\varphi_j^{(s)}$, we can shown that it causes an infidelity $\delta F\approx 2[\varphi_i^{(s)2}+\varphi_j^{(s)2}]/5$. Imbalance between $\varphi_i^{(m)}$ and $\varphi_j^{(m)}$ should also be small: numerically we find that the infidelity scales as $[\varphi_i^{(m)}-\varphi_j^{(m)}]^2$, thus we need $|\varphi_i^{(m)}-\varphi_j^{(m)}|<\pi/100$ for a gate fidelity higher than $99.9\%$.

\subsection{Approximations in the Formulae}
\label{sec:error}
In order to get the analytical expressions for the gate fidelity [Eqs.~(\ref{eq:fidelity},\ref{eq:fidelity_neg},\ref{eq:fidelity_approx})], we have made several approximations in the derivation. Some of them are covered along the derivation in Sec.~\ref{sec:method}: for example, to get the effective Hamiltonian Eq.~(\ref{eq:H_Raman}), we have applied RWA and the adiabatic elimination of the excited state. Their influence can be estimated to be $|\delta|/\omega_{01}$, $|\Omega_{1(2)}|/\omega_{1(2)}$ and $|\Omega_{1(2)}|^2/\Delta^2$. With the Raman transition detuned close to the motional sideband, the first term is about $10^{-4}$ while the other terms are orders of magnitude smaller. Below we address the effects of other approximations.

\emph{Micro-motion}.
For the linear Paul trap, alternating potential is only applied in the $x$ and $y$ directions. Hence there is no micro-motion in the $z$ direction. To the first order in trap parameter $q$, the effect of the transverse micro-motion is to replace $a_k$ with $a_k [1-(q/2)\cos \omega_{\mathrm{rf}} t]/(1-q/2)$ when quantizing the transverse modes \cite{Leibfried2003,Landa2012}, where $q\approx2\sqrt{2}\omega_x/\omega_{\mathrm{rf}} \sim 0.1$ for the parameters we consider. The factor of $1/(1-q/2)$ can be absorbed into the amplitude of each normal mode and therefore only slightly changes the Lamb-Dicke parameter. This can be fully compensated when calibrating the laser intensity. The effect of the RF-frequency term is $O(\eta_k q \Omega_j/\omega_{\mathrm{rf}})$ on $\alpha_j^k$ and $O(\eta_k^2 q\Omega_j/\omega_{\mathrm{rf}})+O(\eta_k^2 q^2\Omega_j/\mu)$ on $\lambda_j^k$ in Eqs.~(\ref{eq:alpha}, \ref{eq:lambda}). Its influence on $\Theta_{jm}$ [Eq.~(\ref{eq:phi})] is more complex. By comparing terms like $\int dt_1\int dt_2\sin\mu t_1\sin\mu t_2\cos\omega_{\mathrm{rf}}t_1 \sin\omega_k(t_1-t_2)$ with the original integration of Eq.~(\ref{eq:phi}), which is $O(1)$, we estimate the error in $\Theta_{jm}$ to be $O(q/\omega_{\mathrm{rf}}\tau)$ and $O(q|\mu-\omega_k|/\omega_{\mathrm{rf}})$. We also numerically evaluate these correction terms for the three gate designs we consider in Sec.~\ref{sec:parameters} and get results lower than these estimations. All of these terms are further squared when calculating fidelity, and the dominate contribution is estimated to be $10^{-6}$ from $\alpha_j^k$. Nevertheless, the correction factor of $[1-(q/2)\cos\omega_{\mathrm{rf}t}]$ can always be incorporated into the formulation if its effect is not negligible.

\emph{Carrier term}.
As is mentioned after Eq.~(\ref{eq:H_expansion}), in the derivation of the XX entangling gate, we dropped a single-qubit rotation term. Strictly speaking this is not an error source because we can apply an additional rotation to compensate it. However, our numerical result shows that for all the three gates we consider in Sec.~\ref{sec:parameters}, such single-qubit rotation terms are less than $10^{-5}$ and can be directly neglected. The reason is that we use multiple laser segments with opposite phases, which largely cancels the single-qubit rotation.

\emph{Higher order terms in the Lamb-Dicke parameter}. In the derivations of Eqs.~(\ref{eq:H_expansion_2},\ref{eq:U_expansion} and \ref{eq:Phi00}-\ref{eq:Phi11}), we only keep zeroth and first order terms in the Lamb-Dicke parameters and the second order diagonal terms with paired $a_k$ and $a_k^\dag$ of the same motional mode. The error in gate fidelity from such approximations is of the order $\eta^4$, and because the Lamb-Dicke parameter $\eta$ always comes with the operators $a_k$ and $a_k^\dag$ whose magnitudes are related to the thermal motion, we express the error as $O(\eta^4(2\bar{n}+1)^2)$, where $\bar{n}$ is the average phonon number of a typical transverse mode. The argument is lengthy and hence is placed in Appendix~\ref{app:higher_order}.

\emph{Asymmetry in blue- and red-detuned coupling.}
 In the derivation of Eq.~(\ref{eq:H_twobeam}), we assume the two pairs of Raman transitions on one ion have the same effective Rabi frequency and opposite detunings (see Fig.~\ref{fig:three_beam}). However, in experiments there are always errors in these parameters, which can significantly influence the gate fidelity. Detailed analyses are placed in Appendix~\ref{app:asymmetry}. The basic idea is that asymmetric detunings lead to a $\varphi^{(s)}$ changing at the rate of $\delta\mu_{\mathrm{asym}}$, hence an error bounded by $O(\delta\mu_{\mathrm{asym}}^2\tau^2)$; while imbalanced effective Rabi frequencies result in an additional rotation in the $y$ direction, and for a gate fidelity higher than 99.9\% we need a relative error less than $0.1\%$ for ion 5 and ion 6, and less than $\epsilon<0.02\%$ for ion 1 and ion 4, and ion 9 and ion 14, using the gate parameters in Sec.~\ref{sec:parameters}. Note that the gate design is robust against global shift of the laser intensity; it is the relative change between the two Raman transition pairs that causes this type of error.

\emph{AC Stark shift}.
The counter-rotating or off-resonant couplings neglected before not only introduce fluctuations between the two qubit states, but also cause a shift in the energy levels, which is known as the AC Stark shift. For ${}^{171}\mathrm{Yb}^+$ ions, the $355\,$nm laser is particularly chosen to reduce the relative shift between the two hyperfine ground states, a.k.a. the differential AC Stark shift. According to Ref.~\cite{Campbell2010}, the differential Stark shift is only about $10^{-4}$ of the effective coupling $\Omega_{\mathrm{eff}}$ between $|0\rangle$ and $|1\rangle$. However, such a relative shift in the energy levels does not correspond to a shift $\delta \mu$ in the Hamiltonian [Eq.~(\ref{eq:H_twobeam})], to which our gate design is not sensitive; in stead
it will increase the asymmetry between the two Raman transition pairs and, as we have mentioned above, will lead to an infidelity of $(\delta \mu_{\mathrm{asym}} \tau)^2$. A constant AC Stark shift can be easily compensated by a corresponding shift in the driving laser's frequencies, but in our case the intensities of the driving laser are also varying. One possible solution is to tune the laser frequencies for each segment accordingly. Another possibility is to use one strong beam and one weak beam for each Raman transition, and only to adjust the weak beam to change the effective coupling strength. For example, in Fig.~\ref{fig:three_beam} we can make the lower beam stronger than the upper one, while still balance the effective Rabi frequencies of the two Raman transitions. By letting the strong beam 10 times as the weak one, we can reduce the changes in AC Stark shift to 1/10 while keeping the effective coupling unchanged.

\emph{Spontaneous emission}.
So far we have not considered the spontaneous emission from the excited state. To couple the two ground states with the off-resonant Raman transition, there is actually a small probability of $\Omega_{1(2)}^2/\Delta^2$ for the ion to be in the excited state, from which the spontaneous emission can occur at the rate of $\gamma_e$. This will lead to decoherence between the two qubit states. For the gate design we consider in Sec.~\ref{sec:parameters}, the error from spontaneous emission is estimated to be $10^{-3}$ for the longest gate time of about $500\,\mu$s, if we set $\Omega_1=\Omega_2$. Note that if we use one strong and one weak beams for Raman transition to reduce the change in AC Stark shift, the spontaneous emission error will be dominated by the stronger beam and hence will be increased.

\subsection{Other Sources of Noise and Errors}
\label{sec:other}
A broad laser beam can cause unwanted transitions on the adjacent ions, while a narrow beam can lead to fluctuation in the laser intensity felt by the target ion due to its thermal motion. Suppose the laser beams have a Gaussian profile, that is, the intensity is proportional to $\exp(-r_\perp^2/2\sigma^2)$ where $\sigma$ is the width of the beam. With $\sigma=2\,\mu$m, when a beam is shined on one ion, its effect on an adjacent ion is of the order $\exp(-d_{\mathrm{av}}^2/2\sigma^2)\sim 10^{-4}$. Meanwhile, the thermal motion perpendicular to the laser beam is dominated by that in the $z$ direction. For a harmonic trap, the error can be estimated to be $(1/32\pi^2)\eta_k^2(2\bar{n}+1)(\lambda/\sigma)^2(\omega_x/\omega_z)^2$, where $\eta_k$ and $\bar{n}$ are for a transverse mode, and $\lambda$ is the laser wavelength. To realize a linear trap along the $z$ direction, we need $\omega_x/\omega_z > 0.77 N/\sqrt{\log N}$ \cite{PhysRevLett.71.2753,Steane1997,Lin2009} for a harmonic trap; the estimation for anharmonic trap is more difficult but the scaling should not be worse. Therefore the noise on the laser intensity due to thermal motion is also of the order $10^{-4}$ for a chain of tens of ions.

The trapping parameters $\omega_{x(y,z)}$ are also subjected to experimental noises. It has mainly two effects: (1) a shift in phonon mode frequencies (phonon mode dephasing), whose effect is roughly the same as an opposite shift in detuning $\mu$ and (2) small change in the equilibrium configuration, whose effect depends on the width of each laser beam. Therefore, our examples of gate design should be able to tolerate $2\pi\times 1\,$kHz shift in the transverse trapping frequencies while still maintains a fidelity of 99.9\%. The weaker axial trapping is achieved by a DC field, hence less vulnerable to fluctuations. For an estimation, we again consider a harmonic potential $\omega_z$. The dimension of length appears as $(q^2/4\pi\epsilon_0 m \omega_z^2)^{1/3}$, hence $\delta z/z \sim -2\delta\omega_z/3\omega_z$. For $N$ ions in the linear chain, the largest change in equilibrium position is for ions on the end, with $z=N d_{\mathrm{av}}/2$. Now if we want the change in the laser intensity to be less than $1\%$ for an ion, i.e. $1-\exp(-\delta z^2/2\sigma^2)\approx \delta z^2/2\sigma^2 \sim 1\%$, we get $\delta\omega_z/\omega_z\sim 0.5\sigma/N d_{\mathrm{av}}\approx 0.5\%$, that is, our gate design can tolerate a relative change of $0.5\%$ in the axial frequency. Actually large shift during one gate time is not very likely; usually trapping parameters vary in a much longer time scale and in principle we can adjust the laser beams before the experiment to compensate such a long-term effect.

We have assumed that the ion chain is sideband cooled to a low temperature before the experiment and stays there. Now we consider the effect of heating in the motional modes, which varies with the trap design. According to Eq.~(\ref{eq:optimization}), our optimization process is not sensitive to the phonon numbers if they stay constant. Hence the infidelity due to the motional heating can be bounded by the ``failure rate'' as $\delta F_{\mathrm{heating}} < N\Gamma\tau$ where $N$ is the number of the transverse modes in use, $\tau$ the gate time and $\Gamma$ the average heating rate.
Currently it is possible to realize a heating rate around 1 phonon/s for our choice of ${}^{171}\mathrm{Yb}^+$ ion and transverse mode frequencies around $2\pi\times 3\,$MHz \cite{Gaebler2016,Turchette2000}. Hence the error is bounded by $\delta F_{\mathrm{heating}} < 10^{-2}$ for $N=19$ and $\tau=500\,\mu$s, but note that this is not a tight bound.
When we conduct a similar numerical simulation as that in Appendix~\ref{app:asymmetry} for two ions and one motional mode, with an additional Lindblad term describing the heating, we find that the infidelity does scale linearly with the heating rate, but the value is about 2 orders of magnitude smaller. Meanwhile, the use of multiple segments should further reduce the error. Therefore we believe that the motional heating is not a dominant source of error for tens of ions.

Another effect not covered is the Kerr coupling, i.e. the dependence of one mode's frequency on the phonon number of another mode due to the nonlinearity in the Coulomb interaction \cite{Roos2008}. It can be calculated by expanding the Coulomb potential to the fourth order.
Numerically we find that the largest coupling is about $0.14\,$Hz/phonon between a transverse mode and an axial mode, and about $0.02\,$Hz/phonon between two transverse modes. The error due to a constant shift in a mode frequency $\omega_k$ can be estimated by that in the detuning $\mu$, because if we apply RWA to Eq.~(\ref{eq:alpha}) and Eq.~(\ref{eq:phi}), only the difference between $\mu$ and $\omega_k$ influences the final fidelity [Eq.~(\ref{eq:fidelity})].
Because our scheme can tolerate such a shift up to $1\,$kHz, we conclude that the Kerr coupling has negligible effects for the examples we are considering.

\subsection{Coherent vs Incoherent Errors}
\label{sec:dnorm}
Up to now we have been using average gate fidelity to evaluate the gate performance, because it is easier to treat theoretically and also easier to measure experimentally through randomized benchmarking. However, it is well-known that high gate fidelity does not immediately imply a low enough error rate, or more precisely the diamond norm, which appears in the statement of the Quantum Threshold Theorem \cite{Sanders2016}. In certain cases these two measures can differ significantly, especially for coherent errors. Since we have motivated our analysis by fault-tolerant quantum computing, it is worthwhile to discuss the relation between our results and the diamond norm.

\begin{figure}[!tbp]
      \centering
      \includegraphics[width=0.9\linewidth]{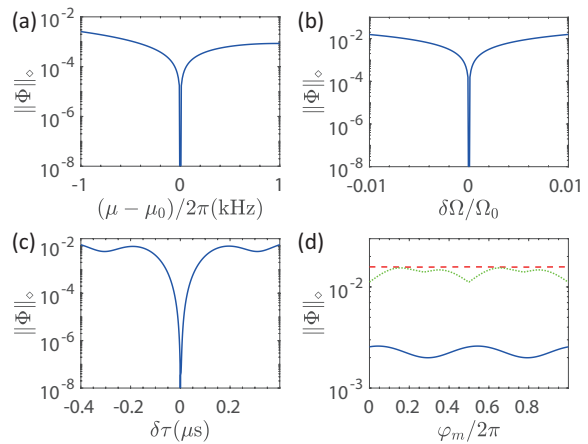}
      \caption{Parameter sensitivity for the entangling gate between ion 5 and ion 6. (a) Diamond norm as a function of shift in detuning. (b) Diamond norm as a function of relative shift in Rabi frequency. (c) Diamond norm as a function of shift in gate time $\tau$. (d) Consider $\varphi_i^{(m)}=\varphi_j^{(m)}$ between 0 and $2\pi$. Solid blue, dashed red and dotted green curves are the maximal diamond norm for a shift of $1\,$kHz in detuning $\mu$, a 1\% change in intensity, and $0.4\,\mu$s change in total gate time, respectively. For (a), (b) and (c), the diamond norms below $10^{-8}$ are not shown, since they are subject to numerical errors.}\label{fig:dnorm_5_6}
    \end{figure}

\begin{figure}[!tbp]
      \centering
      \includegraphics[width=0.9\linewidth]{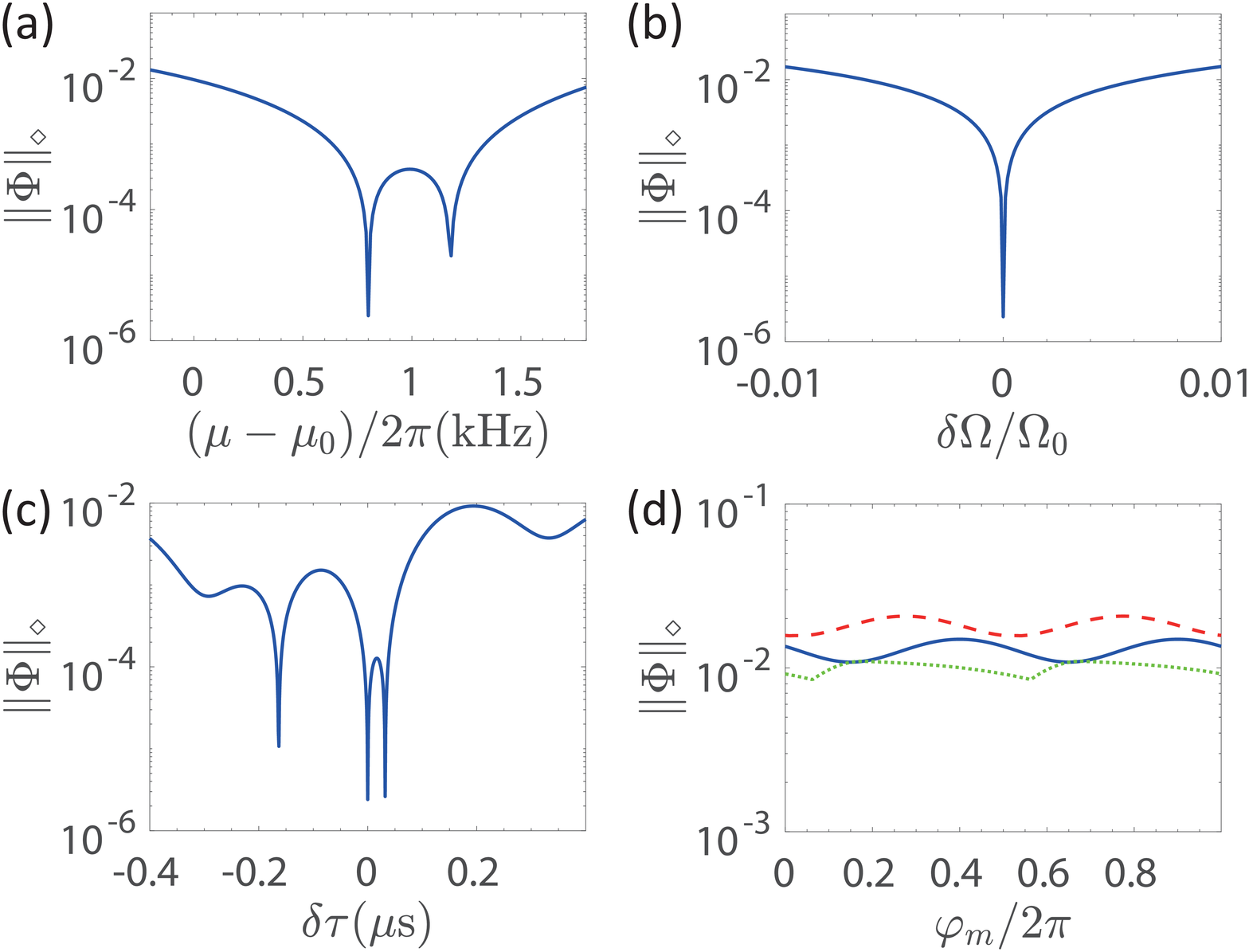}
      \caption{Parameter sensitivity for the entangling gate between ion 1 and ion 4. (a) Diamond norm as a function of shift in detuning. (b) Diamond norm as a function of relative shift in Rabi frequency. (c) Diamond norm as a function of shift in gate time $\tau$. (d) Consider $\varphi_i^{(m)}=\varphi_j^{(m)}$ between 0 and $2\pi$. Solid blue, dashed red and dotted green curves are the maximal diamond norm for a shift of $1\,$kHz in detuning $\mu$, a 1\% change in intensity, and $0.4\,\mu$s change in total gate time, respectively.}\label{fig:dnorm_1_4}
    \end{figure}

\begin{figure}[!tbp]
      \centering
      \includegraphics[width=0.9\linewidth]{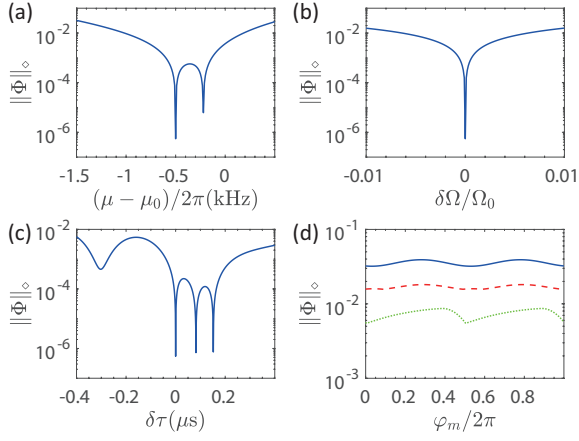}
      \caption{Parameter sensitivity for the entangling gate between ion 9 and ion 14. (a) Diamond norm as a function of shift in detuning. (b) Diamond norm as a function of relative shift in Rabi frequency. (c) Diamond norm as a function of shift in gate time $\tau$. (d) Consider $\varphi_i^{(m)}=\varphi_j^{(m)}$ between 0 and $2\pi$. Solid blue, dashed red and dotted green curves are the maximal diamond norm for a shift of $1\,$kHz in detuning $\mu$, a 1\% change in intensity, and $0.4\,\mu$s change in total gate time, respectively.}\label{fig:dnorm_9_14}
    \end{figure}

The parametric shifts that we considered in Sec.~\ref{sec:parameters} are generally unitary errors. For this type of error it is known that the diamond norm $D$ scales as $\sqrt{\delta F}$ \cite{Kueng2016}, hence our criteria of $\delta F<10^{-3}$ will give a diamond norm of about $3\times 10^{-2}$ in the worst case. Actually as we have already derived the expression of the density matrix [Eq.~(\ref{eq:rho})], we can numerically evaluate the diamond norm using semidefinite programming \cite{Watrous2009,matlab-diamond-norm}. The results for the gate designs we considered in Sec.~\ref{sec:parameters} are shown in Figs.~\ref{fig:dnorm_5_6}, \ref{fig:dnorm_1_4} and \ref{fig:dnorm_9_14}. As expected, the diamond norms at the extreme shifts reach the order of $O(10^{-2})$ and are slightly above the threshold of the surface code of about $1\%$, although for the small system we are considering, it is more meaningful to compare with the pseudothreshold, which is about $8\times 10^{-4}$ for the Surface-17 code \cite{Tomita2014}. This suggests that better controls on the parameters are needed for low enough error rates. Similarly, we expect the bounds on the asymmetry of the beam configurations and the mismatched laser phases to be tighter, as they are also coherent errors. Note however that diamond norm is usually a pessimistic estimation of the errors, and 99.9\% fidelity is nevertheless a good target in practice for many near-term applications of quantum computation and quantum information.

Another observation from these figures is that the error from the imperfect gate design, when there is no parametric shifts, has similar values measured by the infidelity and the diamond norm. This suggests that this gate design error is incoherent and will scale linearly with the number of gates. We discuss this in further detail in Appendix~\ref{app:accumulation}.

The error sources we considered in Sec.~\ref{sec:error} and Sec.~\ref{sec:other} can be divided into classes. The effects of the micro-motion, RWA and the higher order terms in Lamb-Dicke parameters should be mainly unitary because the approximations are directly made in the Hamiltonian. They are estimated to be $O(10^{-4})$ or less, hence the diamond norm should be of order $O(10^{-2})$. On the other hand, spontaneous emission and other errors related to the thermal motions should be incoherent and the diamond norm will not be very different from the infidelity \cite{Kueng2016}.
\subsection{Summary and Discussion}
We summarize in Table~\ref{tab:restriction} the requirements on the relevant parameters for a gate fidelity higher than 99.9\%,
\begin{table}[htbp]
\centering
\caption{Restriction on fluctuations or errors in physical parameters for $F>99.9\%$.
\label{tab:restriction}}
\begin{tabular}{|c|c|}
\hline
Source of Error & Requirement\\
\hline
Slow fluctuation in $\mu$ & $|\delta \mu|<1\,$kHz\\
\hline
Slow fluctuation in $\Omega$ & $|\delta\Omega/\Omega_0|<1\%$\\
\hline
Error in gate time $\tau$ & $|\delta\tau|<0.4\,\mu$s\\
\hline
\tabincell{c}{Detuning asymmetry in \\phase-insensitive setup} & $|\delta\mu_{\mathrm{asym}}|<10\,$Hz\\
\hline
\tabincell{c}{Rabi frequency asymmetry \\in phase-insensitive setup} & $|\delta\Omega_{\mathrm{asym}}/\Omega_0|<0.02\%$\\
\hline
\tabincell{c}{Phase asymmetry \\on the two ions} & $|\varphi_{\mathrm{asym}}^{(m)}|<\pi/100$\\
\hline
Nonzero $\varphi^{(s)}$ & $|\varphi^{(s)}|<\pi/100$\\
\hline
\tabincell{c}{Laser's phase fluctuation\\during one gate time} & $|\delta\varphi|<\pi/100$\\
\hline
\tabincell{c}{Change in trapping\\frequencies} & \tabincell{c}{$|\delta\omega_x| < 1\,$kHz\\ $|\delta\omega_z/\omega_z|<0.5\%$}\\
\hline
\end{tabular}
\end{table}
\begin{table}[htbp]
\caption{Errors from neglected terms and effects. In this table, $q$ is a parameter of the Mathieu equation describing the Paul trap, see e.g. \cite{Leibfried2003}; $\gamma_e$ the spontaneous emission rate of the excited state $|e\rangle$; $\Omega_1$ and $\Omega_2$ the Rabi frequencies corresponding to any a Raman transition pair and $\Omega_{\mathrm{eff}}$ the resulting effective coupling between $|0\rangle$ and $|1\rangle$; $d_{\mathrm{av}}$ the average ion spacing; $\sigma$ the width of each laser beam. See the main text for the definition of other symbols. The last column gives an estimation of the order of magnitude based on the chosen experimental parameters.\label{tab:error}}
\begin{tabular}{|c|c|c|}
\hline
Source of Error & Expression & Value\\
\hline
Micro-motion & $(\eta q \Omega_{\mathrm{eff}}/\omega_{\mathrm{rf}})^2$ & $10^{-6}$\\
\hline
RWA & $|\delta|/\omega_{01}$, $|\Omega_{1(2)}|/\omega_{1(2)}$ & $10^{-4}$\\
\hline
\tabincell{c}{Adiabatic elimination \\of the excited state} & $|\Omega_{1(2)}|^2/\Delta^2$ & $10^{-7}$\\
\hline
\tabincell{c}{Spontaneous emission \\of the excited state} & $\gamma_e \tau |\Omega_{1(2)}|^2/\Delta^2$ & $10^{-3}$\\
\hline
\tabincell{c}{Higher order terms in \\ Lamb-Dicke parameter $\eta$} & $\eta^4(2\bar{n}+1)^2$ & $10^{-4}$\\
\hline
\tabincell{c}{Laser beams on \\ adjacent ions} & $\exp(-d_{\mathrm{av}}^2/2\sigma^2)$ & $10^{-4}$\\
\hline
\tabincell{c}{Thermal motions \\ perpendicular to \\the laser beams} & $\frac{\eta^2(2\bar{n}+1)}{32\pi^2} \left(\frac{\lambda}{\sigma}\right)^2 \left(\frac{\omega_x}{\omega_z}\right)^2$ & $10^{-4}$\\
\hline
\end{tabular}
\end{table}
and in Table~\ref{tab:error} the error from terms and effects neglected in the derivation. As we can see, the most prominent technical challenge in realizing a high-fidelity entangling gate is to compensate any imbalance in the two Raman transition pairs to couple the qubit states; they require very careful control in the frequencies, intensities and beam profiles of the laser.

In comparison, errors from spontaneous emission of the excited state, which is right at the order of 0.1\% in our examples of gate design, seems to set an ultimate limitation: we note that the error from spontaneous emission is proportional to the effective Rabi frequency $\boldsymbol{\Omega}$ (if two beams of the Raman transition have the same intensity) and the total gate time $\tau$. If one is reduced, the other should be increased to realize the desired entangling gate and therefore the error does not decrease. However, what we presented in Sec.~\ref{sec:parameters} are not the shortest possible gate time and weakest possible Rabi frequency: when optimizing parameters such as $\boldsymbol{\Omega}$ and $\tau$, we have considered the robustness against fluctuation in parameters. If these fluctuations can be further suppressed, we can use other solutions with shorter gate time and weaker laser intensity \cite{1710.01378}. Then the infidelity from spontaneous emission can be reduced.

Finally, for the current example we are considering with a few tens of ions, effects of thermal motion and heating are not dominant; but as these effects scale with the ion number $N$, we will need better way to cool the ion chain and to isolate it from the environment when we proceed to larger scale ion trap quantum computing.

\section{Conclusion}
\label{sec:conclusion}
To sum up, in this paper we thoroughly examine the use of transverse motion in a 1D ion chain and segmented optical pulses to realize XX entangling gates. We first review the derivation of the Hamiltonian and the time evolution operator using a phase-insensitive geometry. An analytical expression of the average gate fidelity is presented, whose optimization is equivalent to an eigenvalue problem. This scheme is directly applied to the 17-ion quantum error correction surface code in a chain of 19 ${}^{171}\mathrm{Yb}^+$ ions for a concrete numerical estimation. After calculating the optimized gate parameters for ion pairs with different separations inside the chain, we list the constraint on the fluctuation or errors in these parameters in order to maintain a gate fidelity higher than 99.9\%. We further analyze the contribution from the approximations we have made during the derivation and from some effects we have ignored. It is shown that for the parameters we use, spontaneous emission from the excited state is a dominant source of error at the order of 0.1\%, and the restriction on the asymmetry of the laser beams is also tight. If fluctuation and errors in the parameters can be suppressed in future experiments, both of these effects can be reduced.
\begin{acknowledgments}
The authors would like to thank K. Brown, C. Monroe, J. Kim, K. Kim, Y. Wan, and J. Zhang for helpful discussions. This work was supported by the IARPA Logic Qubit program.
\end{acknowledgments}
\appendix
\section{Equilibrium Positions and Transverse Modes}
\label{app:anharmonic}
For typical experimental parameters in trapped-ion quantum computing, the ions' micro-motion is small. Here we ignore the micro-motion and effectively treat the trap as a static pseudo-potential. An estimation for the introduced error can be found in Sec.~\ref{sec:error}. Suppose strong trappings are applied along the $x$ and $y$ directions. Then ions' equilibrium configuration must be along the $z$ axis. Consider the following potential energy
\begin{equation}
 U= \sum_i \left(-\frac{1}{2}\alpha_2 z_i^2 + \frac{1}{4} \alpha_4 z_i^4\right) + \sum_{i<j}\frac{e^2}{4\pi\epsilon_0 \left|z_i - z_j\right|}
\end{equation}
with $\alpha_2,\alpha_4>0$. By defining the length unit $l_0 \equiv (e^2/4\pi\epsilon_0 \alpha_2)^{1/3}$, dimensionless coordinate $u_i \equiv z_i/l_0$ and dimensionless potential energy $V \equiv 4\pi \epsilon_0 l_0 U/e^2$, we get
\begin{equation}
V= \sum_i \left(-\frac{1}{2}u_i^2 + \frac{1}{4} \gamma_4 u_i^4\right) + \frac{1}{2} \sum_{i\neq j} \frac{1}{\left|u_i - u_j\right|},
\end{equation}
where $\gamma_4 \equiv \alpha_4 l_0^2/\alpha_2$ is a dimensionless constant which completely determines the shape of the equilibrium configuration. For a given number of ions and $\gamma_4$, we can minimize the potential energy to find the equilibrium positions using Newton's method. The gradient and the Hessian matrix of the potential energy can be calculated as
\begin{equation}
\frac{\partial V}{\partial u_m} = -u_m + \gamma_4 u_m^3 - \sum_{j \ne m} \frac{u_m - u_j}{\left|u_m - u_j\right|^3},
\end{equation}
\begin{equation}
\frac{\partial^2 V}{\partial u_m^2} = -1 + 3 \gamma_4 u_m^2 + \sum_{j\neq m}\frac{2}{\left|u_m - u_j\right|^3},
\end{equation}
\begin{equation}
\frac{\partial^2 V}{\partial u_m \partial u_n} = -\frac{2}{\left|u_m - u_n\right|^3} \quad\quad (m \neq n).
\end{equation}

\begin{figure}[!tbp]
  \centering
  \includegraphics[width=\linewidth]{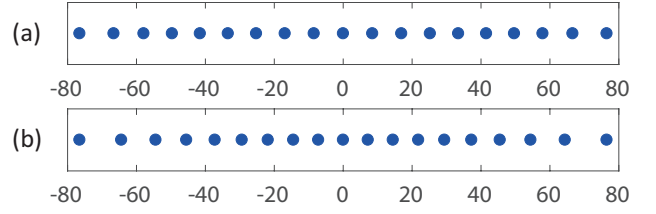}\\
  \caption{(a) Equilibrium positions for an anharmonic trap potential with $\gamma_4=4.3$ and $l_0=40$ (arbitrary unit). (b) Equilibrium positions for a harmonic trap potential that can produce the same average ion spacing.}\label{fig:position}
\end{figure}

For the example we use in Sec.~\ref{sec:result} (17 ions for computation and 2 auxiliary ions at the ends), we adjust $\gamma_4$ to minimize the relative standard deviation (RSD) for the spacings of the 17 computing ions. $\gamma_4=4.3$ is found to give a minimal RSD of only 2.3\%; in comparison, a harmonic trap gives rise to an RSD of 11.2\% (see Fig.~\ref{fig:position}).

After finding the equilibrium positions \{$u_i^{(0)}$\} along the axial $z$ direction, we further consider the normal modes by expanding the complete expression of the potential energy
\begin{align}
U=&\sum_i \left(-\frac{1}{2}\alpha_2 z_i^2 + \frac{1}{4}\alpha_4 z_i^4 + \frac{1}{2}m\omega_x^2 x_i^2 + \frac{1}{2}m\omega_y^2 y_i^2\right) \nonumber\\
& + \frac{e^2}{4\pi\epsilon_0}\sum_{i<j}\frac{1}{|\boldsymbol{r}_i - \boldsymbol{r}_j|}
\end{align}
around the equilibrium positions $x_i^{(0)}=y_i^{(0)}=0$, $z_i^{(0)} = l_0 u_i^{(0)}$. The Taylor series up to the second order is given by:
\begin{align}
U &=U_0 + \frac{1}{2}\sum_{\alpha,\beta,i,j}\frac{\partial^2 U}{\partial r_{\alpha,i}\partial r_{\beta,j}}\Bigg|_{r_{\alpha,i}=r_{\alpha,i}^{(0)}}\nonumber\\
&\qquad\times (r_{\alpha,i}-r_{\alpha,i}^{(0)}) (r_{\beta,j}-r_{\beta,j}^{(0)})+\cdots
\end{align}
where $\alpha,\beta=1,2,3$ correspond to the three Cartesian coordinates, while $i,j=1,2,\cdots,N$ correspond to each ion. Since we are only interested in small oscillations around the equilibrium configuration, the quadratic term in the expansion suffices to describe the motion, which is separable in the $x$, $y$ and $z$ directions. The nonlinear effect of the higher order interaction is briefly discussed in Sec.~\ref{sec:other}. Here we only consider the transverse motion in the $x$ direction, while the modes in the $y$ and $z$ directions can be obtained in a similar way.

Define $z_{ij}\equiv |z_i^{(0)} - z_j^{(0)}|$. At the equilibrium positions we have
\begin{equation}
\frac{\partial^2 U}{\partial x_m^2} = m\omega_x^2 - \frac{e^2}{4\pi\epsilon_0} \sum_{j\neq m} \frac{1}{z_{mj}^3},
\end{equation}
\begin{equation}
\frac{\partial^2 U}{\partial x_m \partial x_n} = \frac{e^2}{4\pi\epsilon_0} \frac{1}{z_{mn}^3} \quad\quad (m \neq n).
\end{equation}
We can then diagonalize this matrix to find the normal modes of the transverse motion, with the $k$-th normalized mode vector denoted by $b_j^k$ ($j=1,2,\cdots,N$). These modes can be quantized to give the phonon Hamiltonian.
\section{Errors from Neglecting Higher Order Terms}
\label{app:higher_order}
In the derivation of Eq.~(\ref{eq:U_expansion}), we keep only zeroth and first order terms in Lamb-Dicke parameters $\eta_k$, as well as the second order terms with balanced creation and annihilation operators such as $\eta_k^2 a_k^\dag a_k$. Here we prove that the neglected higher order terms lead to an error of $O(\eta_k^4)$ in the gate fidelity.

These higher order terms can be divided into four classes:
\begin{enumerate}
\item $O(\eta_k^2)$ off-resonant terms with unpaired $a_k$ or $a_k^\dag$ and a single-qubit rotation, such as
    \begin{equation}
    \eta_k^2 a_k^2 \sigma_j^n \int dt \Omega_j e^{-2i\omega_k t}\cos(\mu t+\varphi_j^{(m)})
    \end{equation}
    and
    \begin{equation}
    \eta_k\eta_l a_k a_l^\dag \sigma_j^n \int dt \Omega_je^{-i(\omega_k-\omega_l)t} \cos(\mu t + \varphi_j^{(m)})
    \end{equation}
    for $k\ne l$, where numerical factors of the order 1 are omitted. These terms have vanishing expectation values in a thermal state and therefore they contribute to the gate fidelity only when pairing with another two creation or annihilation operators. Hence the error is $O(\eta_k^4)$. Also notice that for the example we consider in Sec.~\ref{sec:result}, the spectrum of the transverse modes is narrow and all the $\omega_k$'s are similar and are also close to $\mu$. Therefore the time integral is of the order $\Omega_j/\omega_k \lesssim 1$.
\item $O(\eta_k^3)$ resonant terms with unpaired $a_k$ or $a_k^\dag$ and a single-qubit rotation, such as
    \begin{align}
    &\eta_k\eta_l\eta_m a_k^\dag a_l^\dag a_m \sigma_j^n \times\nonumber\\
    &\int dt \Omega_j e^{i(\omega_k+\omega_l-\omega_m)t}\sin(\mu t+\varphi_j^{(m)}).
    \end{align}
We consider two possible cases here. (i) Two frequencies are the same, e.g. $\omega_l=\omega_m$. Then the time integration has exactly the same form as Eq.~(\ref{eq:alpha}). According to Eq.~(\ref{eq:fidelity_approx}), as we optimize the gate fidelity to higher than 99.9\%, each $|\alpha_j^k|^2$ term should be of the order $10^{-3}$ or less. Besides, the term we drop here has an additional $\eta_k^2$ coefficient compared with $\alpha_j^k$. (ii) All the three frequencies are different. For a wide spectrum, such terms become off-resonant and can be treated in a similar way as $O(\eta_k^2)$ terms; for a narrow spectrum we consider in Sec.~\ref{sec:result}, that is, 19 transverse modes located within a width of about 0.9\% of $\omega_x$, such a term has a shifted $\omega_k$ compared with Eq.~(\ref{eq:alpha}), hence its contribution should be similar to a shifted detuning $\mu$ by the same amount, with the additional $\eta_k^2$ factor. To sum up, such terms have negligible effects so long as the gate fidelity calculated by Eq.~(\ref{eq:fidelity_approx}) is high at the optimized parameters and is robust against shift in detuning. To balance the creation and annihilation operators when taking the trace with a thermal state, this type of terms can be paired with the first order terms as well, which, however, vanish for the optimized parameters.
\item $O(\eta_k^3)$ off-resonant terms with unpaired $a_k$ or $a_k^\dag$ and a two-qubit operation, such as the time integral of the commutator between one first order term and one second order term. To balance the creation and annihilation operators, another $O(\eta_k)$ term must be added. So the final contribution is $O(\eta_k^4)$.
\item $O(\eta_k^4)$ resonant terms with a two-qubit operation, such as the time integral of the commutator between two second order terms, or that between one first order term and one third order term. Such terms lead to a relative error of $O(\eta_k^2)$ in $\Theta_{ij}$ of Eq.~(\ref{eq:phi}). Whether the creation and annihilation operators are balanced or not, an error of $O(\eta_k^4)$ in the fidelity is resulted, because $\Theta_{ij}$ is set to be $\pi/4$ in Eq.~(\ref{eq:fidelity}) and the error only appears as a quadratic term.
\end{enumerate}
Since only two ions appear in the Hamiltonian [Eq.~(\ref{eq:H_expansion_2})] and $(\sigma_j^n)^2=I$, there are no multi-qubit operation terms. There are also terms purely in the subspace of motional states, e.g. the commutator between two first order terms with the single-qubit rotation on the same ion. However, such terms act as a global phase on the subspace of the ions' internal states and are irrelevant to the gate fidelity.

Similar arguments also apply to Eqs.~(\ref{eq:Phi00}-\ref{eq:Phi11}): $O(\eta_k^2)$ terms with unpaired creation or annihilation operators or $O(\eta_k^3)$ terms cause an error of $O(\eta_k^4)$ in fidelity and therefore can be neglected. In the derivation of these equations, we use the Zassenhaus Formula \cite{Magnus1954}:
\begin{equation}
e^{X+Y}=e^X e^Y e^{-\frac{1}{2}[X,Y]}\times \cdots
\end{equation}
After dropping the commutators, which are higher order terms based on the argument above, we obtain Eqs.~(\ref{eq:Phi00}-\ref{eq:Phi11}).
\section{Errors from Asymmetry in Phase-Insensitive Geometry}
\label{app:asymmetry}
As is mentioned in Sec.~\ref{sec:error}, the gate design strongly depends on the symmetry of detunings and effective Rabi frequencies of the two Raman transition pairs. Here we discuss the influence of small asymmetry in the phase-insensitive geometry. A similar analysis can also be applied to the phase-sensitive one.

Suppose one Raman transition pair has effective Rabi frequency $\Omega_j+\delta\Omega_j$ and detuning $\mu+\delta\mu$, while the other pair has effective Rabi frequency $\Omega_j-\delta\Omega_j$ and detuning $-\mu+\delta\mu$. Following similar steps as in Eq.~(\ref{eq:H_twobeam}), we get the effective interaction-picture Hamiltonian
\begin{widetext}
\begin{align}
H_I =& \hbar \Omega_j \cos \left[\mu t + \varphi_j^{(m)} - \Delta k \cdot x_j(t)\right] \left[ \sigma_j^x \cos \left(\delta \mu \cdot t + \varphi_j^{(s)}\right) - \sigma_j^y \sin \left(\delta \mu \cdot t + \varphi_j^{(s)}\right)\right]\nonumber\\
& - \hbar \delta\Omega_j \sin \left[\mu t + \varphi_j^{(m)} - \Delta k \cdot x_j(t)\right] \left[ \sigma_j^x \sin \left(\delta \mu \cdot t + \varphi_j^{(s)}\right) + \sigma_j^y \cos \left(\delta \mu \cdot t + \varphi_j^{(s)}\right)\right].
\end{align}
\end{widetext}
Clearly the effect of $\delta\mu$ is a slow change in the rotation axis and its effect (for small $\delta\mu\cdot\tau$) can be bounded by that of a constant error in $\varphi_j^{(s)}$. So the error from asymmetric detuning is $O(\delta\mu^2\tau^2)$ where $\tau$ is the gate time.

The $\delta \Omega_j$ term corresponds to a rotation in the orthogonal direction, which oscillates at the same frequency as the leading order term but with a phase difference of $\pi/2$. It is more difficult to bound its effect. So instead we tackle this problem numerically. A multiple-phonon-mode problem is still hard to solve, even for a relatively small cutoff of phonon numbers; but for only two ions and one phonon mode, the system can be easily solved by standard numerical integration methods, using a Hamiltonian analogous to Eq.~(\ref{eq:H_Raman}) with two pairs of Raman transitions on each of the ion. Then the result can be compared with the method we used in Sec.~\ref{sec:method}. Since our purpose is just to estimate the order of magnitude, we choose a special initial state $|00\rangle\langle00|\otimes\rho_{\mathrm{th}}$ to calculate fidelity.

For symmetric detunings and effective Rabi frequencies, the result is consistent with the error analysis we make in Sec.~\ref{sec:result} and Appendix~\ref{app:higher_order}. For a nonzero $\delta\Omega_j$, it turns out that the dominant source of error is the additional rotation due to the carrier term of the Hamiltonian, i.e.,
\begin{equation}
H_I^{\mathrm{carrier}} = \hbar \Omega_j \sigma_j^x \cos \mu t - \hbar \delta\Omega_j \sigma_j^y \sin\mu t,\label{eq:H_carrier}
\end{equation}
where we choose $\varphi_j^{(m)}=\varphi_j^{(s)}=0$ and $\delta\mu=0$ for simplicity. Originally the carrier term almost vanishes for the optimized gate parameters; but now with the $\delta\Omega_j$ term, the carrier term leads to an additional small rotation, which causes errors in the final entangling gate.

With this observation, we can now estimate the influence of asymmetric effective Rabi frequencies in the gates we considered in Sec.~\ref{sec:result}. All we need to do is to numerically solve the unitary evolution operator corresponding to the carrier term of the Hamiltonian [Eq.~(\ref{eq:H_carrier})], given the pulse sequence $\boldsymbol{\Omega}$ from Sec.~\ref{sec:parameters}. It must be a single-qubit rotation, and the rotation angle $\delta\phi$ indicates that the error in gate fidelity is of the order $\delta\phi^2$. Supposing the intensities of different laser beams are proportional, we have $\delta\Omega(t)=\epsilon\Omega(t)$ where $\epsilon$ is a small parameter. Considering the gate parameters used in Sec.~\ref{sec:result}, we need $\epsilon<0.1\%$ for ion 5 and ion 6, $\epsilon<0.02\%$ for ion 1 and ion 4, and ion 9 and ion 14, to achieve a gate fidelity higher than 99.9\%.

Finally, we also use this method to estimate the effect of the asymmetry in $\Delta k$, which is about the ratio of the hyperfine splitting to the laser frequency, as is discussed in Sec.~\ref{sec:method}. It turns out that this error is negligible for the parameters we choose.

\section{Accumulation of Gate Design Errors under Repeated Operations}
\label{app:accumulation}
As mentioned in Sec.~\ref{sec:dnorm}, a comparison between the infidelity and the diamond norm suggests that the imperfect gate design leads to an incoherent error. Here we further study the accumulation of such errors when applying the entangling gates sequentially. In our formalism, the effect of multiple gates on the same pair of two ions can be easily modelled as a longer pulse sequence. Suppose we have $m$ gates starting at $T_1,\,T_2,\cdots,\,T_m$ respectively, with $T_1<T_1+\tau\le T_2<T_2+\tau\le\cdots\le T_m < T_m+\tau$ where $\tau$ is the gate time. Then Eq.~(\ref{eq:alpha}) and Eq.~(\ref{eq:phi_simple}) should be modified to
\begin{align}
\label{eq:alpha_m_gates}
\alpha_j^k =& -\frac{i}{\hbar} \eta_k b_j^k \int_{T_1}^{T_1+\tau} \chi_j(t) e^{i\omega_k t} dt\nonumber\\
& -\frac{i}{\hbar} \eta_k b_j^k \int_{T_2}^{T_2+\tau} \chi_j(t) e^{i\omega_k t} dt\nonumber\\
& -\cdots\nonumber\\
& -\frac{i}{\hbar} \eta_k b_j^k \int_{T_m}^{T_m+\tau} \chi_j(t) e^{i\omega_k t} dt,
\end{align}
and
\begin{align}
\label{eq:Theta_m_gates}
\Theta_{ij}=&\frac{2}{\hbar^2}\sum_k \eta_k^2 b_i^k b_j^k \Bigg\{\bigg(\int_{T_1}^{T_1+\tau} dt_1 \int_{T_1}^{t_1}dt_2
\nonumber\\
&+\int_{T_2}^{T_2+\tau}dt_1\int_{T_1}^{T_1+\tau}dt_2+\int_{T_2}^{T_2+\tau} dt_1\int_{T_2}^{t_1}dt_2\nonumber\\
&+\cdots\nonumber\\
&+\int_{T_m}^{T_m+\tau} dt_1\int_{T_1}^{T_1+\tau}dt_2+\cdots\nonumber\\
&+\int_{T_m}^{T_m+\tau} dt_1\int_{T_m}^{t_1}dt_2 \bigg)\chi_i(t_1) \chi_j(t_2)\times\nonumber\\
&\sin \left[\omega_k (t_1-t_2)\right]\Bigg\}.
\end{align}

\begin{figure}[!tbp]
\centering
\includegraphics[width=0.9\linewidth]{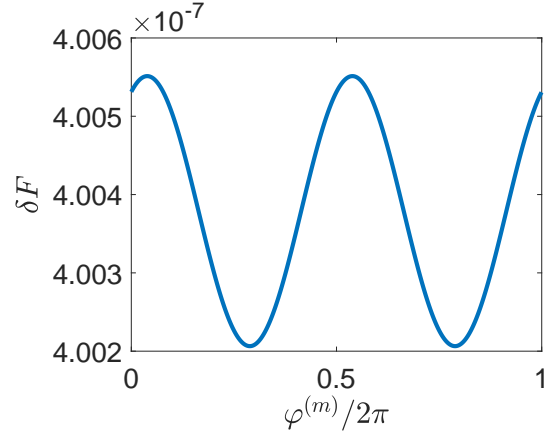}
\caption{Infidelity due to residual coupling to the phonon modes vs. motional phase $\varphi^{(m)}$. The pulse sequence is optimized for $\varphi^{(m)}=0$, but the infidelity is almost independent of $\varphi^{(m)}$.}\label{fig:alpha_motional_phase}
\end{figure}

Changing the starting point of the gate will introduce a random phase factor to each term of $\alpha_j^k$. This ensures that the accumulated infidelity of $m$ gates, which is proportional to $\sum_{jk}|\alpha_j^k|^2$ [see Eq.~(\ref{eq:fidelity_approx})], will scale as $m$ instead of $m^2$. Besides, the varying starting point of the integration in $\alpha_j^k$ also leads to a varying motional phase $\varphi_j^{(m)}$ for each gate. In Sec.~\ref{sec:parameters} we optimize the gate parameters at $\varphi_j^{(m)}=0$ for both ions, but later in Fig.~\ref{fig:alpha_motional_phase} we will show that these gate designs are robust for a nonzero motional phase.

Now we consider the $\Theta_{ij}$ term. First note that any double integrations involving two gates will vanish in Eq.~(\ref{eq:Theta_m_gates}), because then these integrations have a similar form as those in $\alpha_j^k$ and are suppressed by our optimization. The error from these terms will further be squared when computing the fidelity, hence can be safely neglected. Then we are left with $m$ double integrations, each corresponding to an individual gate. Here a random motional phase will also appear due the the varying starting points of the gates. Ideally each double integration should be $\pm\pi/4$ and the total phase $\pm m\pi/4$, but the random motional phase will cause a distribution of the integral. To suppress the accumulated error, we set the mean of this distribution at $\pm\pi/4$, assuming a uniform distribution of the motional phase over $[0,2\pi)$. Then the deviation of the sum of the $m$ gates from $\pm m\pi/4$ will be $O(\sqrt{m})$ and therefore the infidelity will scale as $m$.

\begin{figure}[!tbp]
\centering
\includegraphics[width=0.9\linewidth]{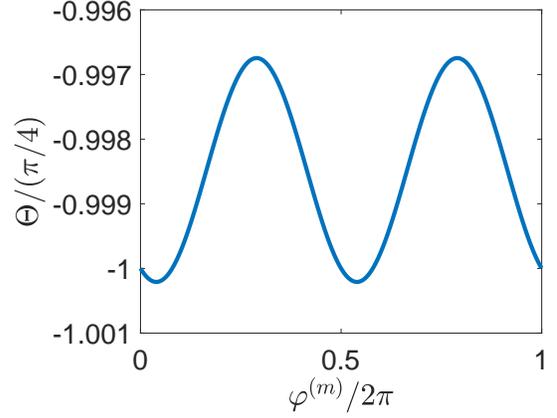}
\caption{$XX$ rotation angle $\Theta_{ij}$ vs. motional phase $\varphi^{(m)}$. The pulse sequence is chosen such that $\Theta_{ij}=\pm\pi/4$ at $\varphi^{(m)}=0$.}\label{fig:Theta_motional_phase}
\end{figure}

As an example, we plot the infidelity due to $\alpha_j^k$ terms (residual coupling to the phonon modes) in Fig.~\ref{fig:alpha_motional_phase} and the value of $\Theta_{ij}$ in Fig.~\ref{fig:Theta_motional_phase} as functions of the motional phase $\varphi_j^{(m)}$. These plots are computed from our gate design for ion 9 and ion 14, where first we determine the shape of the pulse sequence at the detuning $\mu_0=0.997\omega_x$, then we move to the working point $\mu_0'=\mu_0-2\pi\times 0.5\,$kHz and rescale the pulse intensity to set $\Theta_{ij}=\pm\pi/4$. These parameters are the same as those used to get Fig.~\ref{fig:ion_9_14}. As we can see, the residual coupling to the phonon modes is very insensitive to the motional phase.

\begin{figure}[!tbp]
\centering
\includegraphics[width=0.9\linewidth]{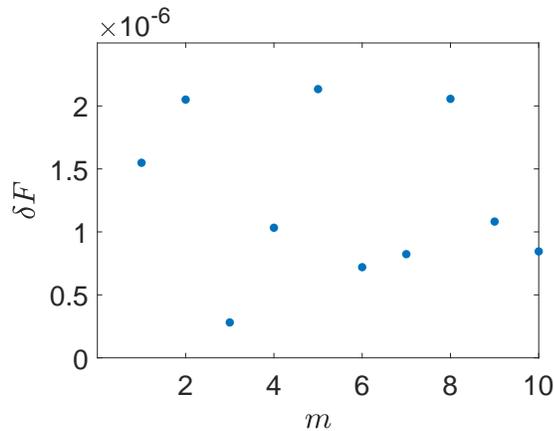}
\caption{Total gate design infidelity vs. number of repeated gates $m$ between ion 9 and ion 14. Here we consider a special case where there is no interval between two adjacent gates.}\label{fig:accumulation}
\end{figure}

Finally we show an example of applying repeated gates by setting $T_i+\tau=T_{i+1}$ ($i=1,\,2,\cdots,\,m-1$). The gate infidelity due to the imperfect design vs. $m$ is plotted in Fig.~\ref{fig:accumulation}. Here we have further rescaled the pulse sequence to move the mean of Fig.~\ref{fig:Theta_motional_phase} to $-1$. No clear accumulation in the gate infidelity is observed, partially because the starting point of each gate is not randomly chosen. Nevertheless, we expect the accumulated error to be $O(m)$ rather than $O(m^2)$.

We should emphasize that with the existence of spontaneous emission, phonon mode dephasing and heating, the above formalism will finally break down as $m\tau$ goes above the coherence time. The effect of these sources of errors are discussed in Sec.~\ref{sec:error} and the purpose of this appendix is just to show that the gate design error is not a dominant source in our scheme even if multiples gates are applied.

\end{document}